\documentclass[reprint,amssymb,amsmath,aip,cha]{revtex4-1}
\usepackage{graphicx}
\usepackage[caption = false]{subfig}
\usepackage{color}
\usepackage{epsfig}
\usepackage{rotating}
\usepackage{ifpdf}
\usepackage{bm}
\usepackage[colorlinks=true,linkcolor=blue]{hyperref}%
\expandafter\ifx\csname package@font\endcsname\relax\else
 \expandafter\expandafter
 \expandafter\usepackage
 \expandafter\expandafter
 \expandafter{\csname package@font\endcsname}%
\fi
\usepackage{cleveref}
\begin{document}
\title{3-Dimensional Dusty Plasma in a Strong Magnetic Field: Observation of Rotating Dust Tori}
\author{Mangilal Choudhary}
\email{Mangilal.Choudhary@exp1.physik.uni-giessen.de} 
\affiliation{I. Physikalisches Institut, Justus--Liebig Universität Giessen, Henrich--Buff--Ring 16, D 35392 Giessen, Germany}
\author{Roman Bergert}
\author{Slobodan Mitic}
\author{Markus H. Thoma}
\begin{abstract} The paper reports on the dynamics of a 3-dimensional dusty plasma in a strong magnetic field. An electrostatic potential well created by a conducting or non-conducting ring in the rf discharge confines the charged dust particles. In the absence of the magnetic field, dust grains exhibit a thermal motion about their equilibrium position. As the magnetic field crosses a threshold value (B $>$ 0.02 T), the edge particles start to rotate and form a vortex in the vertical plane. At the same time, the central region particles either exhibit thermal motion or $\vec{E} \times  \vec{B}$  motion in the horizontal plane. At B $>$ 0.15 T, the central region dust grains start to rotate in the opposite direction resulting in a pair of counter-rotating vortices in the vertical plane. The characteristics of the vortex pair change with increasing the strength of the magnetic field (B $\sim$ 0.8 T). At B $>$ 0.8 T, dust grains exhibit very complex motion in the rotating torus. The angular frequency variation of rotating particles indicates a differential or sheared dust rotation in a vortex. The angular frequency increases with increasing the magnetic field from 0.05 T to 0.8 T. The ion drag force and dust charge gradient along with the E-field are considered as possible energy sources for driving the edge vortex flow and central region vortex motion, respectively. The directions of rotation also confirm the different energy sources responsible for the vortex motion.  
\end{abstract}
\maketitle
\section{Introduction}
One of the characteristic features of a dusty plasma is the strong Coulomb interaction among the nearby negatively charged dust particles. The interaction strength determines its dynamical behavior to the external forces. Therefore, the investigation of the response of the dust grain medium to the external forces such as electrostatic force, neutral drag force, radiation pressure force, thermophoretic force, Lorentz force is of great interest \cite{shukladustybook}. The external forces either affect the motion of dust grains directly or through the dynamics of the plasma species. External magnetic fields play a double role to control the dynamics of the dust grains, namely first through the Lorentz force acting on the plasma species (electrons and ions) and secondly through the Lorentz force acting on the dust particles. The motion of electrons and ions in the external magnetic field (B) affects the dust particles via the Coulomb collisions among them, which has been proven theoretically as well as experimentally \cite{kawrotation,magnetirotationecrplasma,dzlievarotationstrongb}. However, the Lorentz force acting on the dust grains in the presence of a magnetic field needs to be explored in more detail. In laboratory magnetized plasma, there are big challenges to keep the plasma homogeneity and to create a stable dust cluster in an electrostatic potential well with a negligible role of $\vec{E} \times  \vec{B}$  ion drift on the dust grains at strong B (B $>$ 3 T), where one can realize the role of B-field to rotate the charged grains due to the Lorentz force. There are challenges to choose an appropriate potential well, selection of an appropriate particle size and types of materials, and discharge conditions to realize the hypothesis of the Lorentz force on the dust grains in a magnetized plasma. 
\par  
In recent years, a lot of effort has been made to study the dust dynamics in the magnetized plasma background. Maemura {\textit {et al.}} \cite{dusttransportinb} reported the transport of negatively charged particles in a DC discharge plasma when the applied magnetic field is perpendicular to the ambipolar electric field (E), which is a result of the $\vec{E} \times  \vec{B}$  drift of the magnetized electrons. The role of the longitudinal magnetic field (along the discharge axis) to the dust cluster confined in the electrostatic trap of the strata in a DC glow discharge has been studied in detail \cite{vasilievdcrotationinb,vasilmagneticrotation,karasevlongitudinalbrotation,
dustrotationindcwithb,dustrotationindcwithb1,dzlievarotationstratamagnetic,
dzlievarotationstrongb,karasevstrongb}. In these studies, they observed the rotation of dust clusters about the discharge axis due to the $\vec{E} \times  \vec{B}$  drift motion of ions. In a different DC discharge configuration, Sato {\textit {et al.}} \cite{satorotation} studied a volumetric or 3D dusty plasma in the presence of a magnetic field. They demonstrated the role of the coupling among the charged particles to establish the rotational motion in the magnetized plasma. Apart from DC discharges, a set of experiments have been performed in the magnetized rf discharges, where the rf sheath provides an electrostatic trap to the charged particles. The experiments of Konopka {\textit{et al.}} \cite{knopkamagneticrotation} and Huang {\textit{ et al.}} \cite{huangdustrotationrf} have demonstrated the role of the magnetic field on a 2D strongly coupled dust cluster, which exhibits a rigid and sheared rotation due to the $\vec{E} \times  \vec{B}$  drift of ions in the azimuthal direction. The dynamics of a pair of grains in a planar 2D structure containing maximum 12 particles in the presence of a magnetic field has been studied by Cheung {\textit{ et al.}} \cite{cheung2to12planarrotation} and Ishihara {\textit{ et al.}} \cite{2dclusterrotation}. The dust cluster rotation is supposed to be based on the  $\vec{E} \times  \vec{B}$  drift of the ions in the azimuthal direction of a cylindrical geometry. The azimuthal motion of ions transfers their momentum to the dust grains \cite{kawrotation,rotationinionflow}. The azimuthal motion of the neutral gas due to the drifted ions may also cause the dust rotation in the presence of a magnetic field \cite{nedospasovrotationtheory1,gasinduceddustroataion}. Instead of rotation, dust grains can also settle in the current filaments and show various kinds of patterns or ordered structures \cite{mirkpatterninb,thomasorderinb} in strong magnetic fields. All these studies were performed for a 2D dust cluster or 2D dusty plasma in the presence of an external magnetic field. Only limited work on the 3D dusty plasma is performed in magnetized DC and rf discharges \cite{satorotation,dynamiccirculation}. Saitou {\textit {et al.}} \cite{satoprl} have observed the dynamic circulation of charged particles in a 3D dusty plasma in the presence of a magnetic field (B $\sim$ 0.2 T). However, there are still many unanswered questions, e.g.,  about the evaluation of vortices with magnetic field. Do such vortices exist even at strong B-field?. What are the possible sources to drive such vortices?. Does a dusty plasma exist at central region at higher B-field (B $>$ 0.2 T)?. Is such motion common in other 3D dusty plasma system in presence of a B-field?. To get the answer of some of these questions, 3D dusty plasma studies in strong magnetic field are required. The aim of the present investigation is to explore the dynamics of 3D dusty plasma in strongly magnetized rf discharge.\par
Section~\ref{sec:exp_setup} deals with the detailed description of the experimental set-up and the plasma and dusty plasma production. The dynamics of a 3D dusty plasma confined by conducting and non-conducting rings is discussed in Section~\ref{sec:dynamics}.  The origin of the observed vortex flow on the basis of available theoretical models is discussed in Section~\ref{sec:discussion}. A brief summary of the work along with concluding remarks is provided in Section~\ref{sec:summary}.

\section{Experimental setup}  \label{sec:exp_setup}
\begin{figure*}
 \centering
\subfloat{{\includegraphics[scale=0.48]{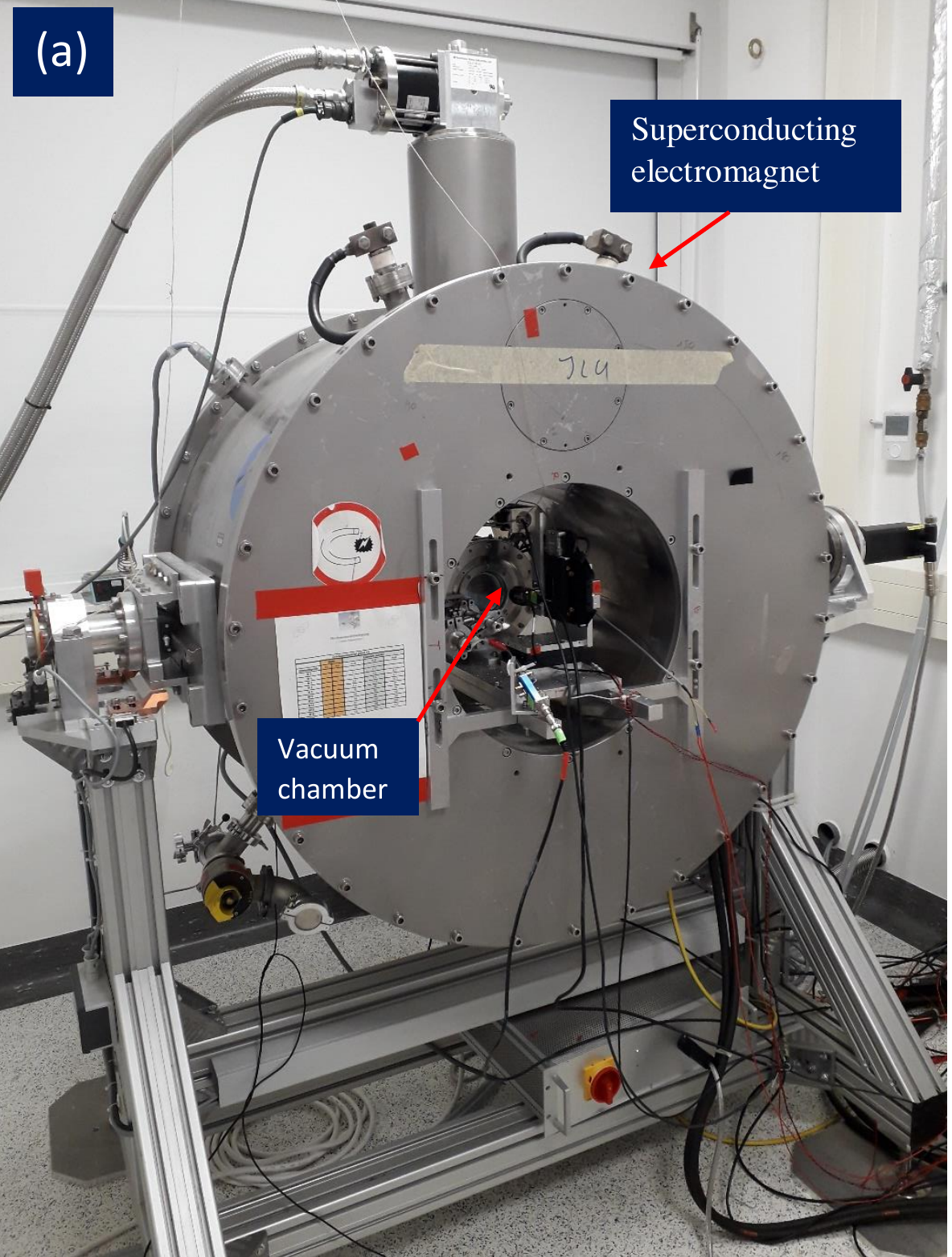}}}%
\qquad
\subfloat{{\includegraphics[scale=0.480]{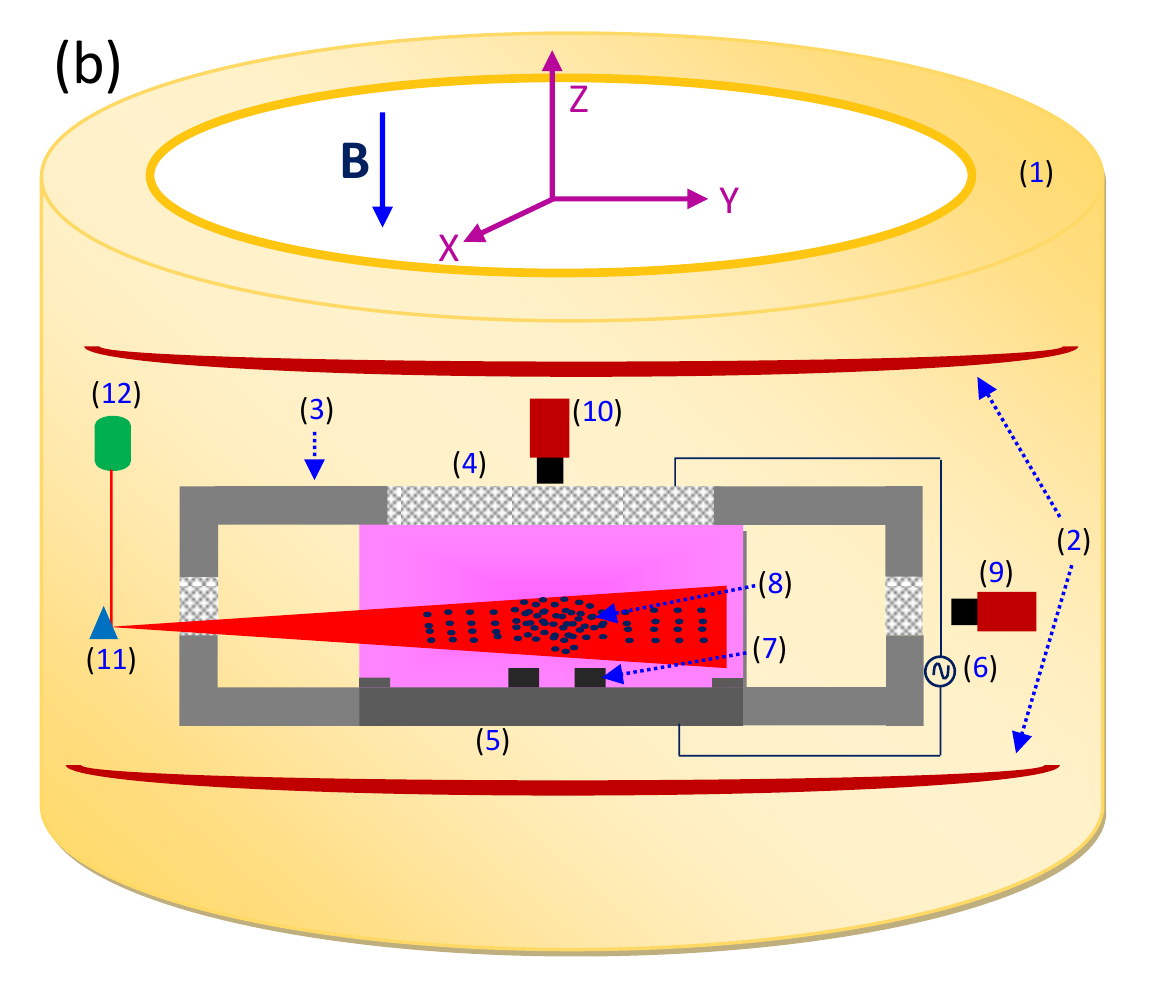}}}
 \caption{\label{fig:fig1} (a) Magnetized Dusty Plasma Device at JLU. (b) Schematic diagram of the experimental setup for the dusty plasma study in a strong magnetic field (1) superconducting electromagnet, (2) magnet coils, (3) vacuum chamber, (4) upper ITO coated transparent electrode, (5) lower aluminium electrode, (6) rf power generator, (7) aluminium or Teflon ring, (8) levitated dust particles, (9) CCD camera for vertical view, (10) CMOS camera for the horizontal view, (11) mirror, and  (12) red laser with a cylindrical lens. The blue arrow indicates the direction of magnetic field (B).}
 \end{figure*}

\begin{figure*}
 \centering
\subfloat{{\includegraphics[scale=0.505]{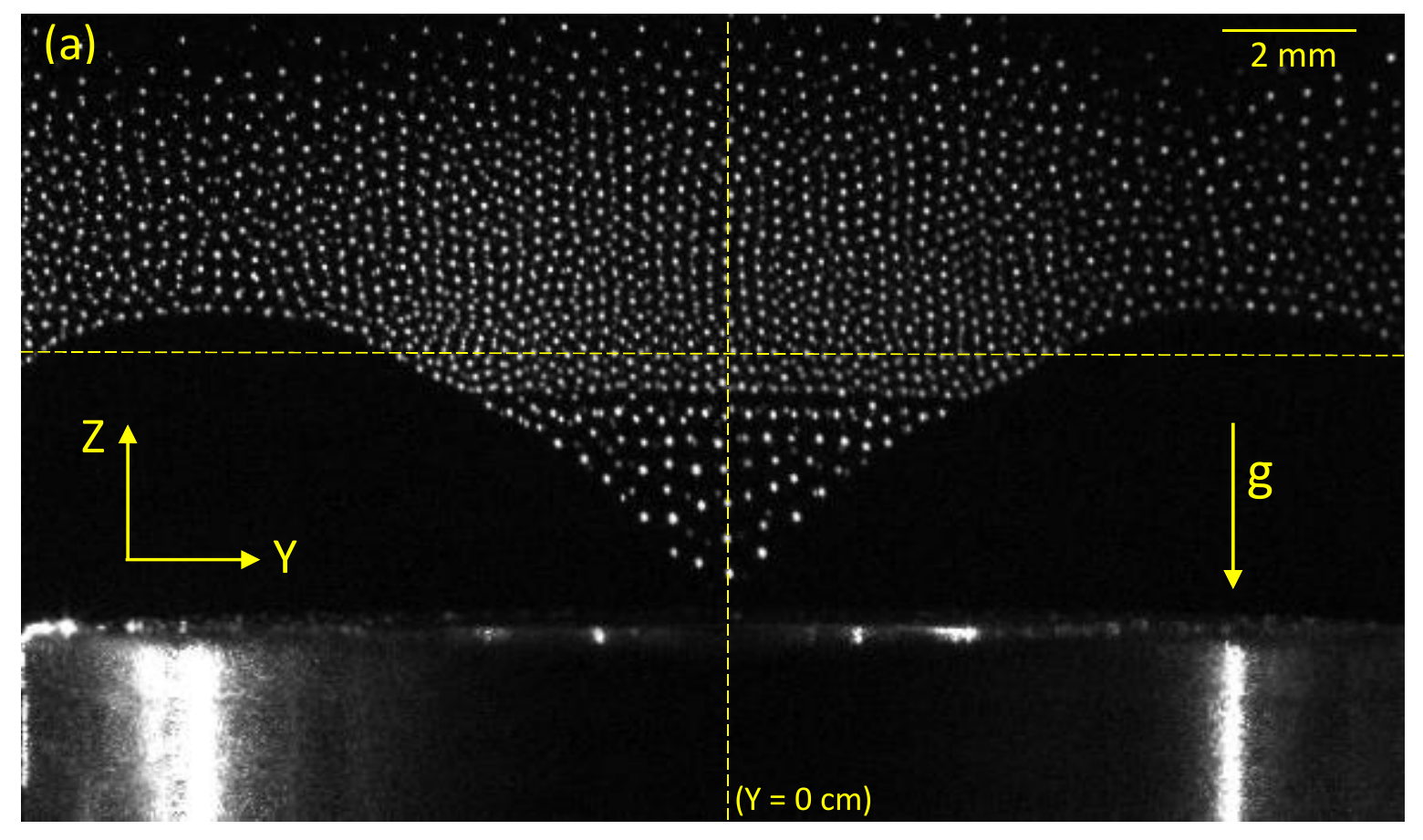}}}%
\qquad
\subfloat{{\includegraphics[scale=0.3850]{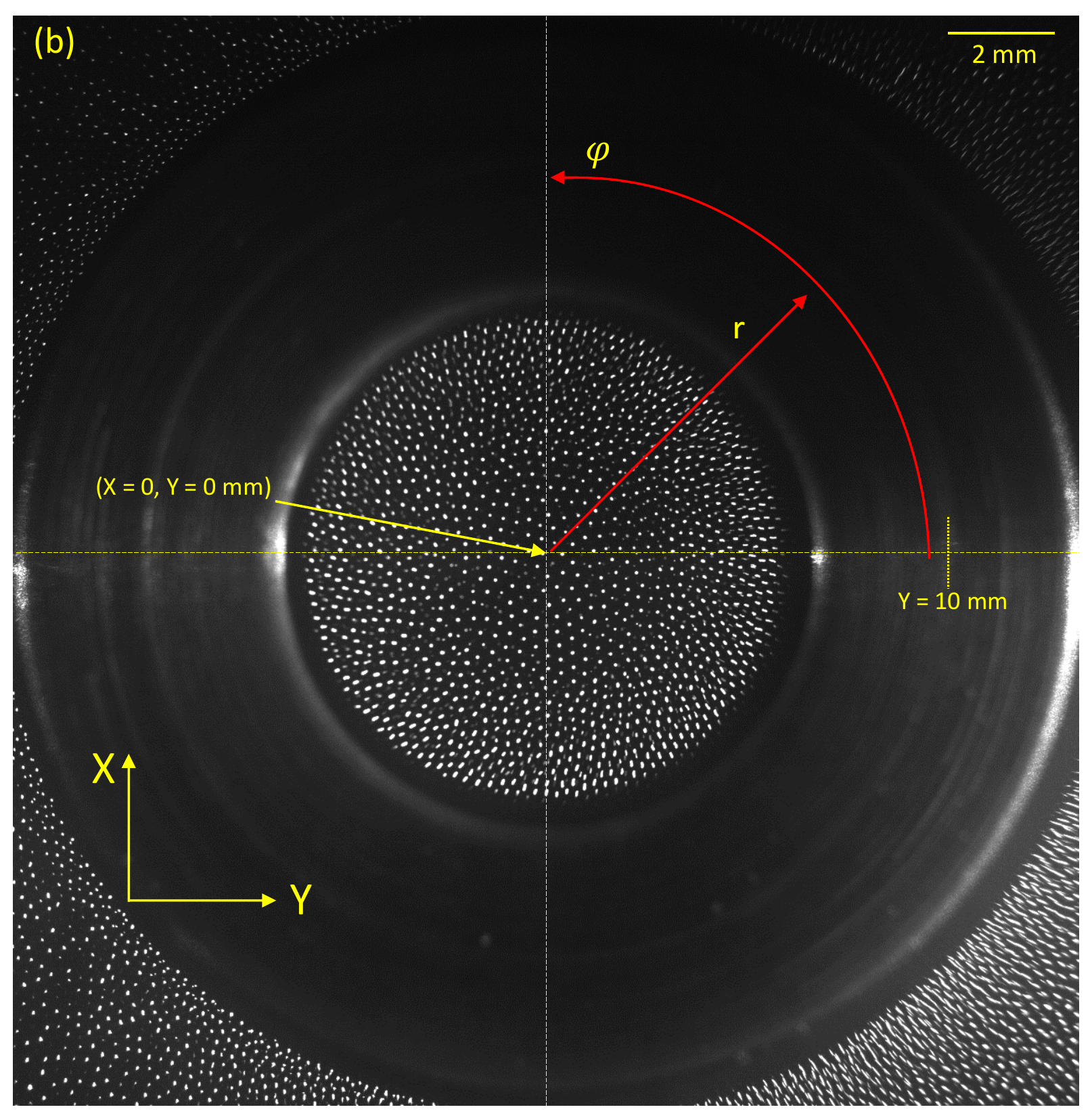}}}
 \caption{\label{fig:fig2} (a) Full view of dusty plasma in a vertical plane (Y--Z) at the centre of the ring (X = 0 cm). A vertical yellow dashed line passes through the centre of Y-axis. The lower electrode surface represents Z = 0 cm. (b) The horizontal view of the dusty plasma at the yellow dashed line (at Z $\sim$ 0.8 cm) in Fig.\ref{fig:fig2}(a)}. 
 \end{figure*}

Experiments are performed in an aluminium vacuum chamber, which is placed at the center of a
superconducting electromagnet ($B_{max}$ $\sim$ 4 T) to introduce a homogeneous magnetic field in the plasma (or dusty plasma). The magnetized dusty plasma device which has been used in the present study is shown in Fig.~\ref{fig:fig1}(a). The orientation of the electromagnet from the horizontal (Fig.~\ref{fig:fig1}(a)) to the vertical (Fig.\ref{fig:fig1}(b)) can be changed. The present experiments are performed in the vertical orientation of the electromagnet. A schematic diagram of the experimental setup is presented in Fig.\ref{fig:fig1}(b). More details about the superconducting electromagnet can be found in ref\cite{mangilalpsst}. At first, the vacuum chamber is evacuated to base pressure p $<$ $10^{-2}$ Pa using a pumping system consisting of a rotary and turbo molecular pump. The experiments are performed with argon gas and the pressure inside the chamber is controlled by using a mass flow controller (MFC) and gate valve controller. For a given argon pressure, the plasma is ignited between an aluminium electrode of 65 mm diameter (lower electrode) and an indium tin oxide (ITO) coated electrode of 65 mm diameter (upper electrode) using a 13.56 MHz rf generator with a matching network. Both electrodes are separated by 30 mm and the upper transparent electrode is grounded along with the vacuum chamber. The special design of the lower electrode, having a ring-shaped periphery with a height of 2 mm and width of 5 mm, provides a radial confinement to the negatively charged dust particles which are levitated above the lower electrode. An additional conducting (aluminium) or non-conducting (Teflon) ring is used to create a 3D dusty plasma in a strong magnetic field. The motivation behind the use of different materials (aluminium and Teflon) for the confinement rings is to observe the role of the depth of the electrostatic potential well  on the dynamics of dust grain medium in an magnetized plasma. It has been observed that dust grains without an additional confining ring or a larger diameter ring are lost from central region to the edge of lower electrode  in the case of strong magnetic field (B $>$ 0.2 T).  
Many rings of different inner diameter (30 mm, 40 mm, and 8 mm) and height (2 mm, 3 mm and 4 mm) are tested for the present study. We have found an appropriate ring of inner diameter 10 mm, outer diameter 20 mm, and height 4 mm to create a deep electrostatic potential well to study the 3D dusty plasma in a strong B. Moreover, particle selection is also necessary to create a 3D dusty plasma at ground level experiments. Our primary measurements with different sized Melamine Formaldehyde (MF) particles suggest to use particles of radius, $r_d$ $\simeq$ 1.7 $\mu$m to create an appropriate volume of dusty plasma at low power (P = 3.5 W). The vacuum chamber has 8 sides (or radial) ports.  A dust dispenser, which is installed at one of the side ports of the vacuum chamber, is used for injecting the dust particles into the plasma. Two opposite side ports are used to observe the dusty plasma using a red laser and a CCD camera in the vertical plane. The CCD camera records images in the vertical plane (Y--Z plane) at 20 fps with a resolution of 1024 $\times$ 768 pixels. A CMOS camera is also installed to observe the dust dynamics through the transparent upper electrode in the horizontal plane (X--Y plane) at a frame rate of 90 fps and with a resolution of 2048$\times$2048 pixels. A full view of the dusty plasma in the vertical (Y--Z) plane at X = 0 cm is shown in Fig.\ref{fig:fig2}(a). Fig.~\ref{fig:fig2}(b) represents the dusty plasma view in the horizontal (X--Y) plane at Z $\sim$ 0.8 cm. The stored images are analyzed with the help of ImageJ \cite{imagejsoftware} software and MATLAB based open-access software, called openPIV \cite{piv}.\par

\section{Dust dynamics in the presence of magnetic field}  \label{sec:dynamics}
The role of an external magnetic field on the dusty plasma has been explored by performing experiments using a conducting (aluminium) and non-conducting (Teflon) ring, which creates a deep and shallow potential well to confine the negatively charged dust particles. There is a stereoscopy technique to track the 3-dimensional motion of particles in the confining potential well \cite{3dvorticesrfplasma}. However, there are limitations in our experiments to diagnose the 3-dimensional dusty plasma using this technique. Therefore, the dynamics of the dust grains in the vertical and horizontal planes are studied to understand the dynamics of 3-dimensional dusty plasma in a strong magnetic field. The detailed results are discussed in the following subsections.
\subsection{Dynamics of the dusty plasma confined by a conducting ring}
\begin{figure*}
\centering
 \includegraphics[scale=0.86]{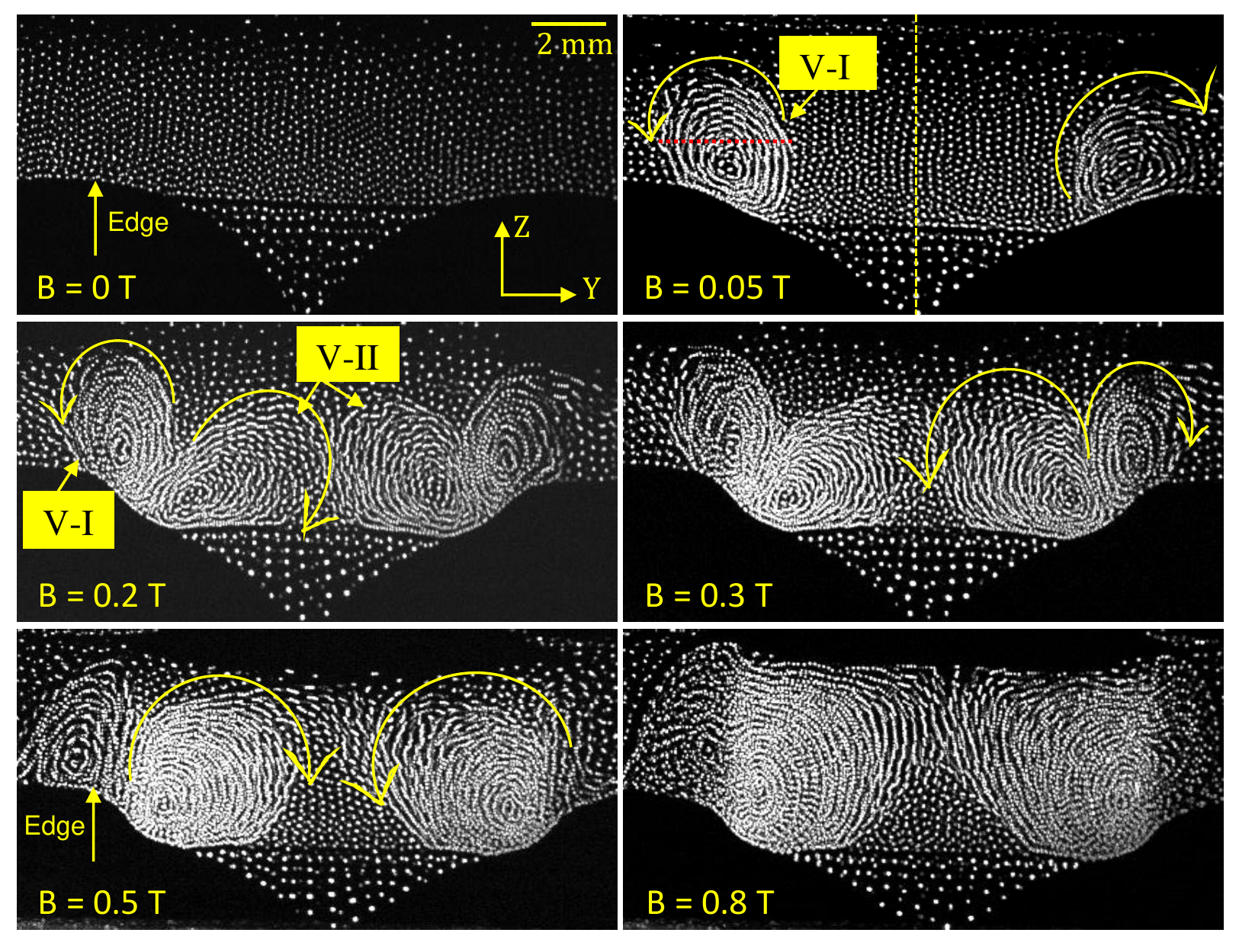}
\caption{\label{fig:fig3} Video images of the dust cloud (aluminium ring) in the vertical (Y--Z) plane at X = 0 cm. Images at different magnetic fields are obtained by a superposition of five consecutive images at a time interval of 65 ms. The edge vortex and central region vortex are represented by V-I and V-II, respectively. The vortex structures at different strengths of the magnetic field are observed at fixed input rf power, P = 3.5 W and argon pressure, p = 35 Pa. The dotted yellow line represents the axis of symmetry. The yellow solid line with an arrow indicates the direction of the vortex flow in the vertical plane of the 3D dusty plasma. The red dash line represents the size (or diameter) of edge vortex.}\par
\end{figure*}

\begin{figure}
\centering
 \includegraphics[scale=0.37]{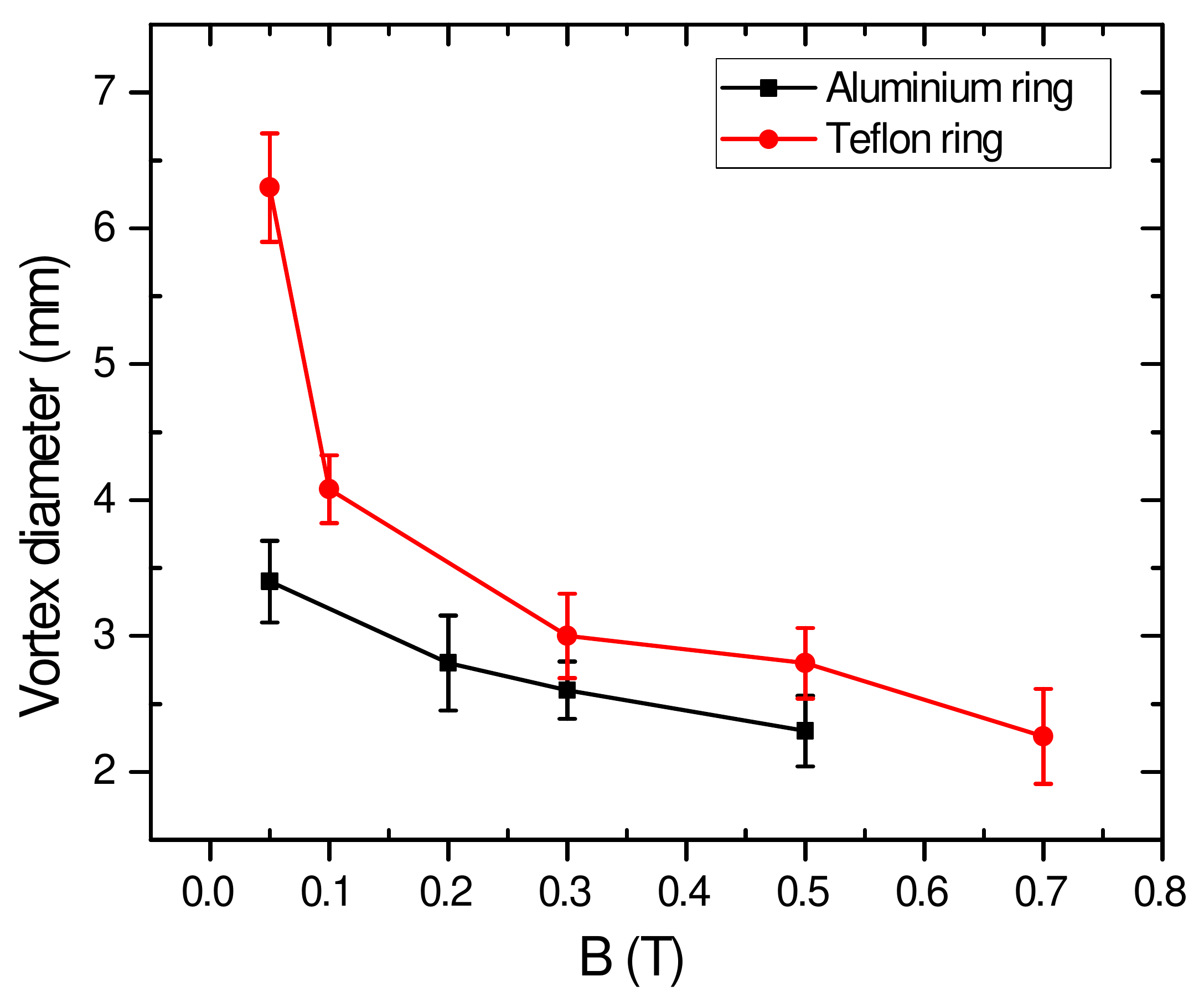}
\caption{\label{fig:fig4} Variation of the size of the edge vortex (V-I) depending on the external magnetic field strength.}
\end{figure}

\begin{figure*}
\centering
 \includegraphics[scale=0.53]{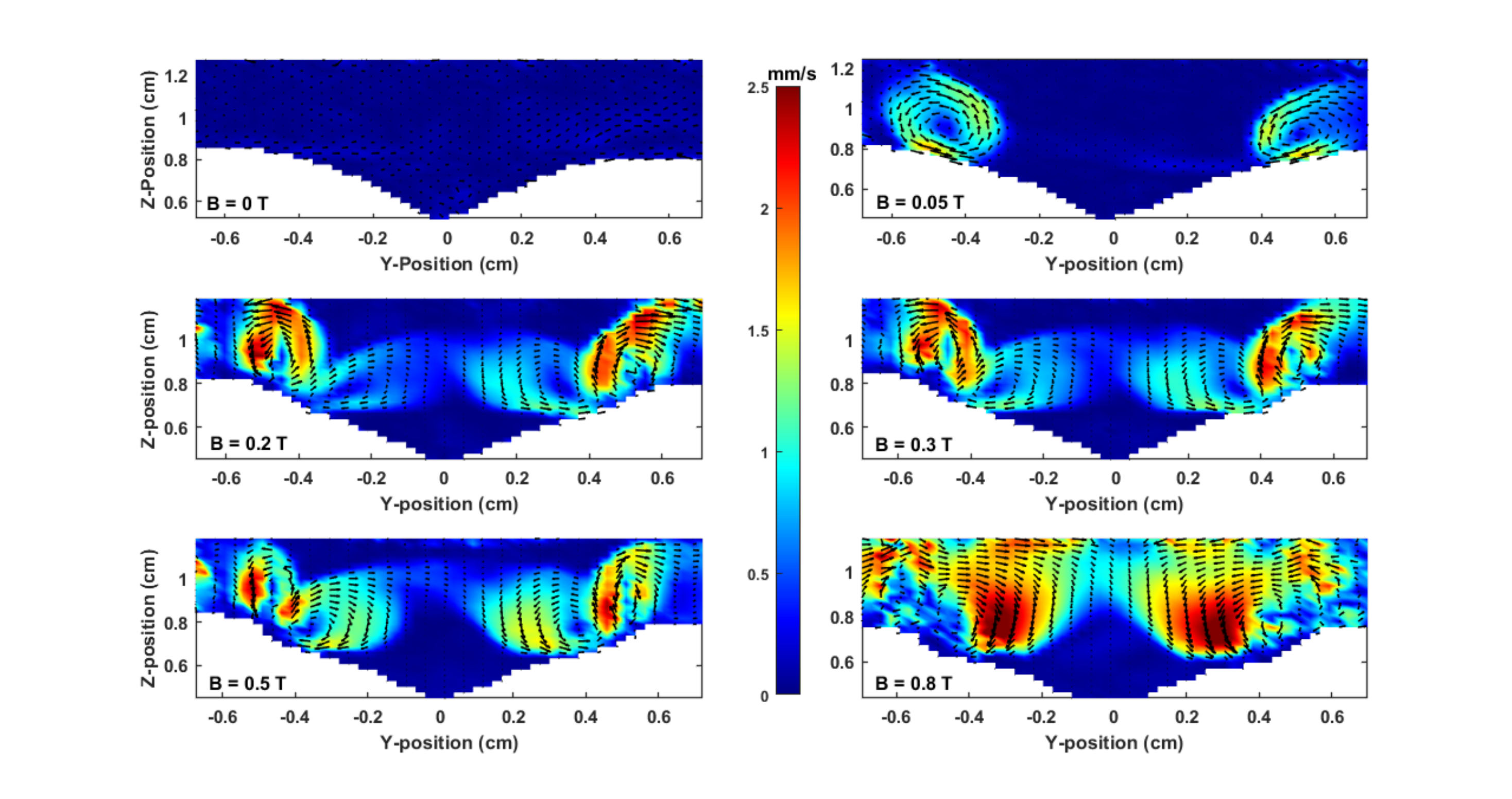}
\caption{\label{fig:fig5} PIV images of the corresponding video images of Fig.~\ref{fig:fig3} at different magnetic field strengths in the vertical (Y--Z) plane. These images are constructed after averaging the velocity vectors over 50 frames. Arrows indicate the direction of rotation in a vortex and color bars represent the magnitude of the velocity of the rotating particles.}
\end{figure*}

An aluminium ring modifies the potential distribution in the sheath region of the lower electrode \cite{rfsheathpotentialdistributation}. It is known that charged particles follow the equipotential contour to achieve an equilibrium position. Therefore, particles are confined in a bowl shaped potential well created by the aluminium ring. Experiments are carried out at an argon pressure of p = 35 Pa and input rf power of P = 3.5 W. It should be noted that gas pressure and input rf power are major discharge parameters causing the plasma current filaments in a strong B-field. In our device, current filaments appear at threshold magnetic field (B $>$ 1) T in the low pressure (p $<$ 10 Pa) and high power (P $>$ 5 W)  discharge. The threshold value of B-field shifts to the higher value with increasing the gas pressure at given rf power. Since the experiments are performed at high pressure (p = 35 Pa) and low rf power (P = 3.5 W),  plasma filaments start to appear above B $>$ 2.2 T. Thus, dust grains are confined in the potential well of the homogeneous plasma in the given magnetic field regime (B $>$ 1 T).
Fig.~\ref{fig:fig3} shows the collective dynamics of 3D dusty plasma in the vertical (Y--Z) plane (at X = 0 cm) at various strengths of the external magnetic field. In this figure, all the images are reconstructed by the superimposition of five consecutive still images. The directed motion of dust grains forms elongated tracks, whereas the thermal motion of grains leaves white dots. In the absence of an external magnetic field, particles exhibit a thermal motion about their equilibrium position. As the magnetic field is increased up to 0.03 T (B $>$ 0.02 T), the edge particles start to rotate in the vertical plane and form a vortex structure (V-I) on either side of the ring edge. Due to the symmetry of dusty plasma about the axis passing through Y = 0 cm in this plane (see Fig.~\ref{fig:fig1}(a)), only the left side of the vortex structure is considered for the further analysis. At low B, the central region particles exhibit either thermal motion or  $\vec{E} \times  \vec{B}$  drift motion in the X--Y plane (or in the azimuthal direction of the cylindrical coordinate), which will be discussed later. Further increase of the magnetic field B $>$ 0.15 T leads to a reduction in the size of the edge vortex (V--I) (see Fig.~\ref{fig:fig4}) and a rotation of the central region particles in the opposite direction, forming a pair of counter-rotating vortices. Since the edge vortices seem nearly circular, the diameter of vortex in the radial (or along Y-axis) direction is considered the size of vortex. At B = 0.2 T, particles in the edge vortex (V--I) and central region vortex (V--II) are observed to rotate in opposite directions and form a counter-rotating vortex pair in this plane. The characteristics (shape, size, angular velocity etc.) of the counter-rotating vortices changes with further increase of the magnetic field up to 0.8 T. At strong magnetic field, B $>$ 0.8 T, the volume of dusty plasma or dust layers in vertical direction decreases. Hence, the size of the central vortex (V--II) reduces and the particles also exhibit a complex vortex motion which is difficult to diagnose with the existing optical diagnostics. It is noticed that the levitation height of the dust cloud shifts from Z $\sim$ 0.6 mm to 0.4 mm (see Fig.~\ref{fig:fig5}), which indicates the reduction of the sheath thickness at strong magnetic field.\par
To investigate the details of the velocity distribution and angular frequency ($\omega$) of rotating particles in a vortex structure the still images are analyzed using a MATLAB based open-access software, called openPIV \cite{piv} software. The PIV images of the dusty plasma in the vertical plane at different strengths of the magnetic field (Fig.~\ref{fig:fig3}) are depicted in Fig.~\ref{fig:fig5}. These PIV images are constructed using an adaptive 2-pass algorithm (a 64$\times$ 64, 50\% overlap followed by a 32 $\times$ 32, 50\% overlap analysis). The contour maps of the average magnitude of the velocities are constructed after averaging the velocity vectors of consecutive 50 frames. In the color map of the PIV images, the direction of the velocity vector represents the rotational direction of the dust grain medium \cite{williamspivanalysis}. Colour bars show the magnitude of the velocity distribution of rotating particles in the vortex structure. It is clear from Fig.~\ref{fig:fig5} that the rotating particles have inhomogeneous velocity distribution in the vortex structure and the velocity of the rotating particles in either the edge vortex (V--I) or the central vortex (V--II) increases as the magnetic field is increased from 0.05 T to 0.8 T. The error over the averaged value of velocity at a given location of vortices (V--I and V--II) at different values of B-field varies between 10 to 20\%, which is estimated by analysing two consecutive PIV images for a set of 50 frames. It is noticed that rotating particles in the vertical plane moves to the next plane. Therefore, dust grain motion can not be considered as an ideal 2D motion in this plane. It clearly indicates the presence of a third velocity component.  It is observed that density of edge particles decreases toward higher B-field. These both factors, presence of third velocity component and less dust density in interrogation area, are mainly possible error sources in PIV analysis \cite{williamspivanalysis}.\par
\begin{figure*}
 \centering
\subfloat{{\includegraphics[scale=0.35]{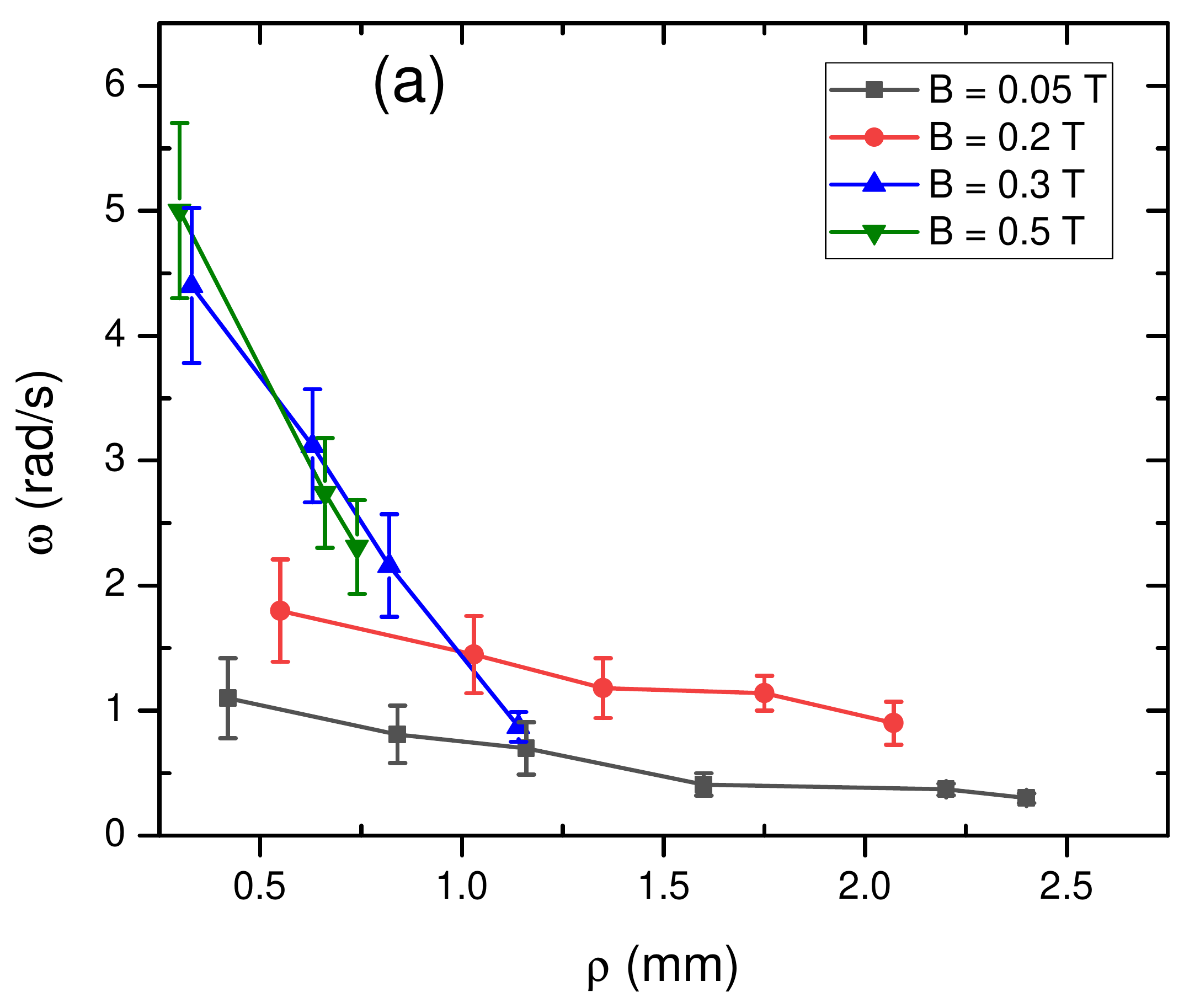}}}%
\qquad
\subfloat{{\includegraphics[scale=0.35]{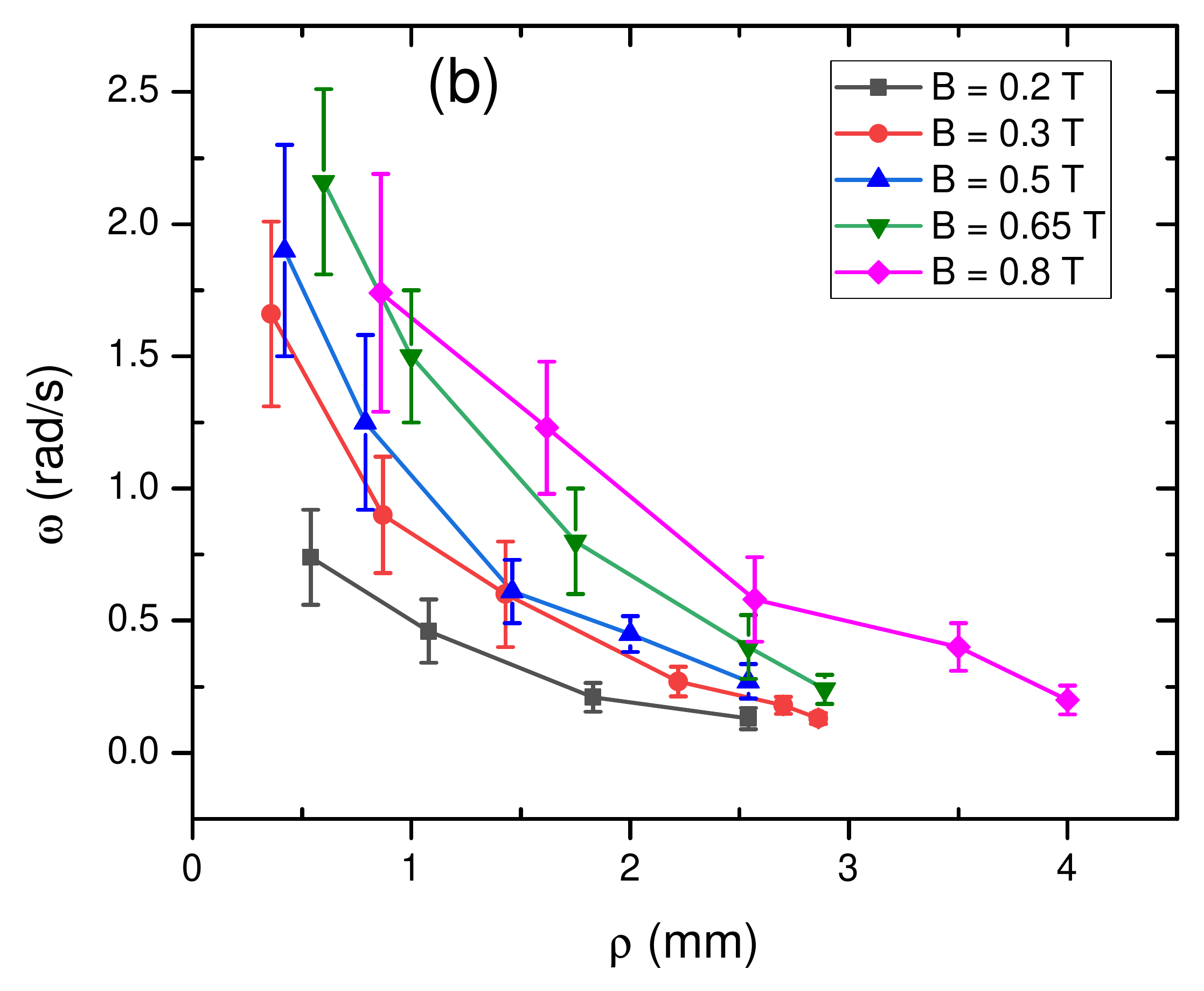}}}
\caption{\label{fig:fig6} The variation of the averaged angular frequency of dust particles in the (a)  edge vortex (V--I)  and (b) central region vortex (V--II) at various strengths of the magnetic field.}
 \end{figure*}

To get the angular frequency, $\omega = v_t/\rho$, of the rotating particles in the edge vortex and central region vortex, circular paths or arcs about the center of the vortex ($\rho$ = 0 mm) are analysed for different values of B-field. Here $v_t$ is the tangential  (or azimuthal) velocity and $\rho$ is the radial distance from the center of a vortex. It is evident that central vortices as well as some of the edge vortices are not circular. Therefore, the tangential velocity is estimated along the circular paths (arcs) at different values of $\rho$. Since $v_t$ is not homogeneous along the circular path, different values of $\omega$ are obtained at given $\rho$. For the comparative study, an average value of $\omega$ is calculated from different values of $\omega$ at given $\rho$. An average angular frequency of the rotating particles in the edge vortex (V--I) and  central region vortex (V-II) is presented in Fig.~\ref{fig:fig6}(a) and Fig.~\ref{fig:fig6}(b), respectively. It is clearly from fig.~\ref{fig:fig6} that the angular frequency of the rotating particles decreases as the radial distance of the particles increases from the center of the vortex, $\rho$ = 0 mm, for a given magnetic field. This is a signature of a differential or sheared rotation of the particles in the vortex. The angular frequency of the rotation increases with increasing external magnetic field at a given radial location from the center of the vortex, as shown in Fig.~\ref{fig:fig6}. It is also clear from figure~\ref{fig:fig6} that the dust particles of the edge vortex (V--I) always have higher values of $\omega$ as compared to the central vortex (V--II) at a given value of the magnetic field.\par
\begin{figure*}
\centering
 \includegraphics[scale=0.55]{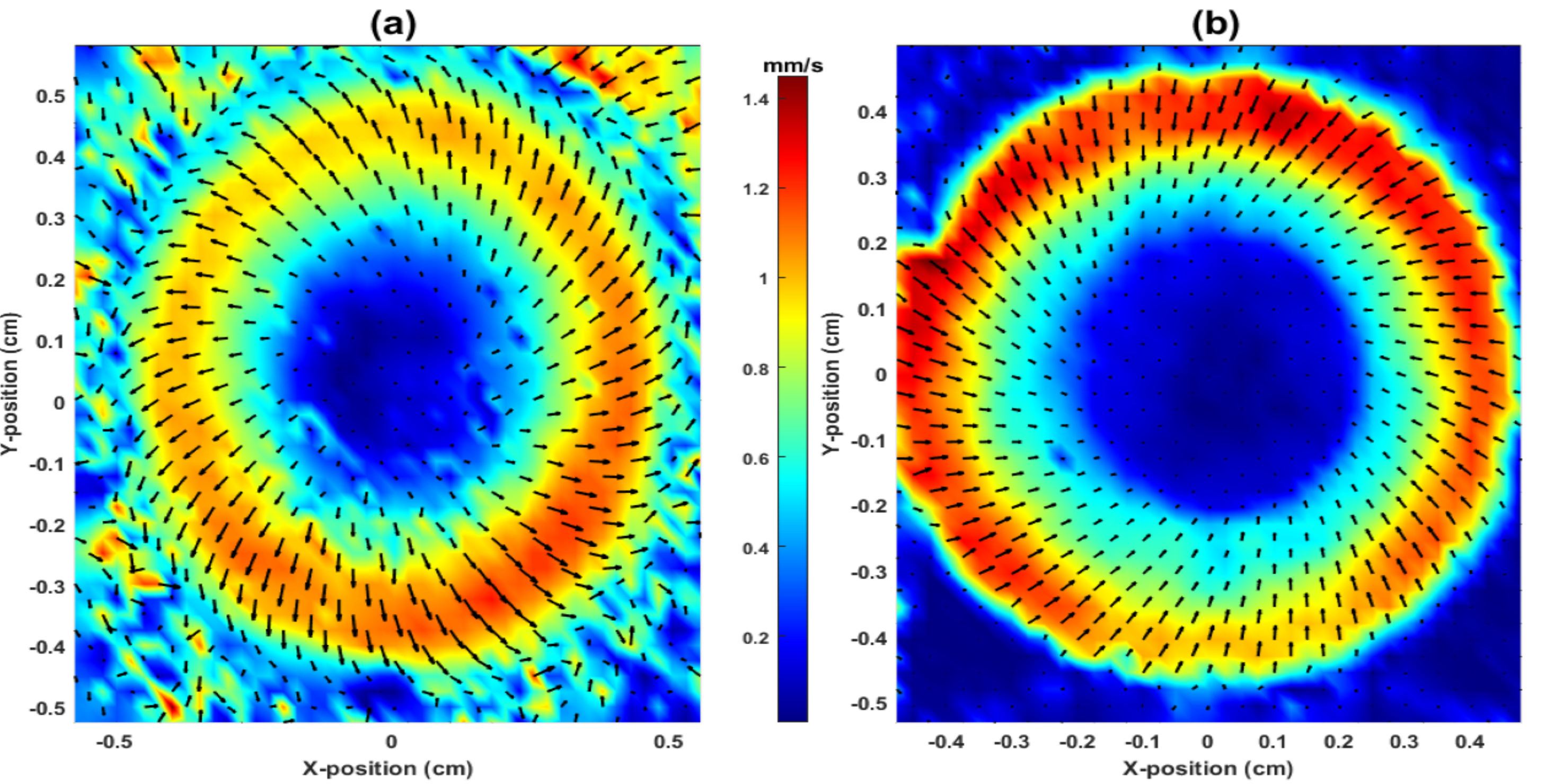}
\caption{\label{fig:fig7} (a) PIV velocity maps, constructed after averaging the velocity vectors over 50 frames in the horizontal (X--Y) plane at (a) Z $\sim$ 1.1 cm and (b) Z $\sim$ 0.7 cm in the presence of an external magnetic field of B = 0.4 T. The potential well is created by using the aluminium ring at argon pressure, p = 35 Pa and input rf power, P = 3.5 W.}
\end{figure*}

For getting more information on the vortex flow in a 3D dusty plasma, images of the horizontal (or X--Y) plane of another experiment at similar discharge conditions are analyzed at different Z-positions. In Fig.~\ref{fig:fig7}, two PIV color map images of the dusty plasma at Z $\sim$ 1.1 cm and Z $\sim$ 0.7 cm in the presence of a magnetic field of B = 0.4 T are displayed. The direction of velocity vectors is outwards in Fig.~\ref{fig:fig7}(a) and inwards in Fig.~\ref{fig:fig7}(b), which corresponds to the edge vortex (V--I) and central region vortex (V--II), respectively. Now it is clear from figure~\ref{fig:fig7} that the edge vortex and central region vortex in the Y--Z plane are cross sections of the rotating dust torus. A single vortex corresponds to a rotating dust torus. At higher magnetic fields, a pair of counter-rotating vortices in the Y-Z plane is nothing but a cross-section of a pair of counter-rotating tori in a dusty plasma. The characteristics of the central region vortex (V--II) (dust torus) is found to be similar to that reported in unmagnetized rf discharge by Mulsow \textit{et al.}\cite{3dvorticesrfplasma}. It is also observed that the velocity vectors have some angle to the radial direction (in the cylindrical coordinate) which suggest an $\vec{E} \times  \vec{B}$ drift motion of the dust particles in the azimuthal direction along with rotational motion in the Y--Z plane. There may also be the possibility to have such kind of motion due to the dust-neutral collisions during the motion in the vertical plane. Therefore, the rotating dust particles in Y--Z plane always have a drift in the azimuthal direction. It clearly indicates that particles do not rotate in a single Y--Z plane or a cross section of the dust torus but they shift to the next plane during the rotation. After a certain value of the magnetic field (B $>$ 0.5 T), the rotating particles have a higher drift velocity (or azimuthal component) which is a signature of the helical type of motion in a dust torus.\par
\begin{figure}
\centering
 \includegraphics[scale=0.45]{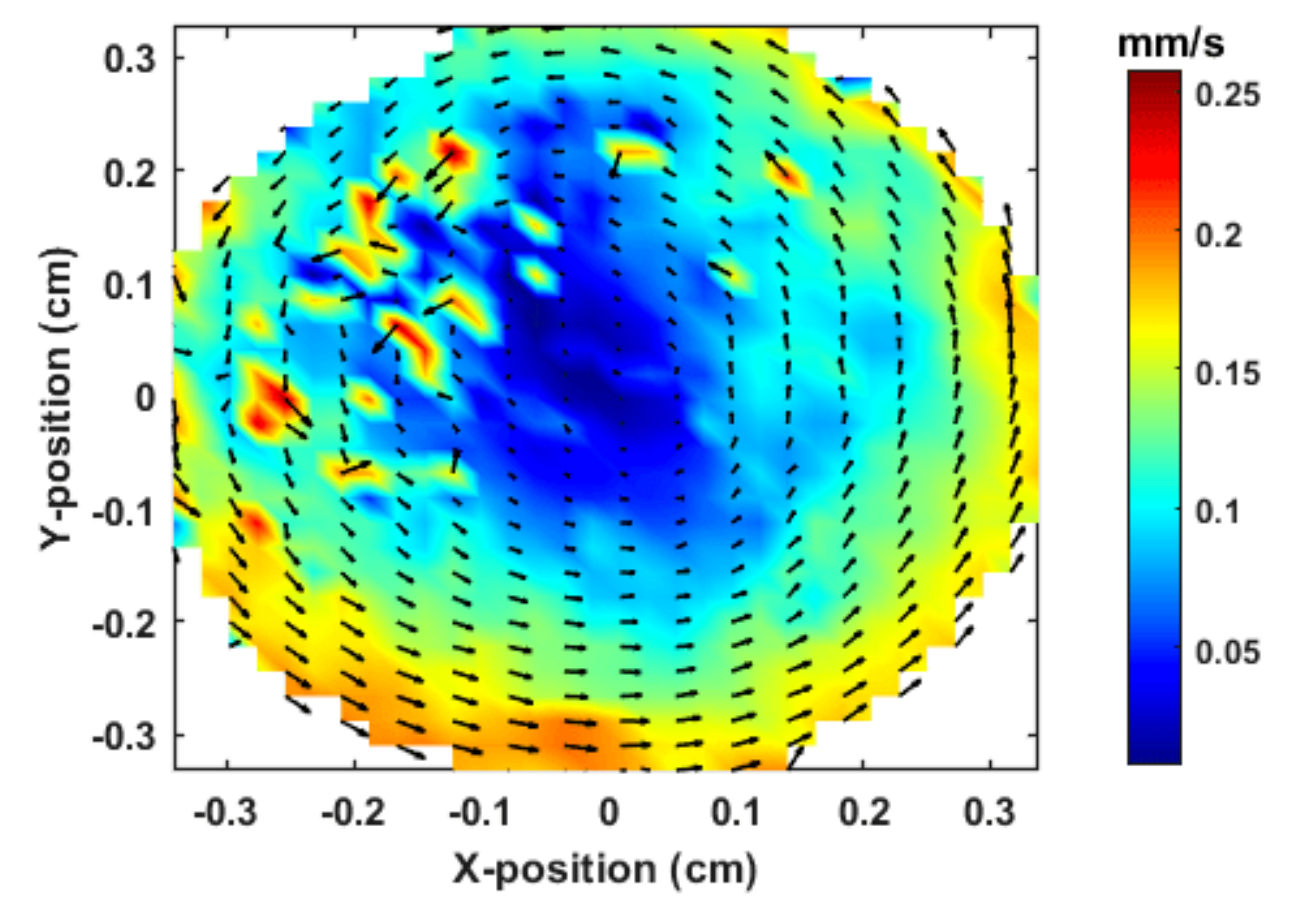}
\caption{\label{fig:fig8} PIV velocity maps, constructed after averaging the velocity vectors over 40 frames in the horizontal (X--Y) plane at Z $\sim$ 0.6 cm in the presence of external magnetic field of B = 0.2 T. The potential well is created by using the aluminium ring at argon pressure, p = 35 Pa and input rf power, P = 3.5 W.}
\end{figure}

 It is well known that the dust-dust interaction strength, which is termed as Coulomb coupling, is required  for the vortex flow. In the case of weak coupling, dust grains do not show a collective response to external or internal forces. In our case, the particles levitated in the strong electric field of the sheath or the deep potential well (see Fig.~\ref{fig:fig2}(b)) seem to be weakly coupled because they do not participate in the vortex flow. These particles are acted upon by the ion drag force in the azimuthal or $\vec{E_r} \times  \vec{B}$  direction and rotate in the horizontal plane, as shown in Fig.~\ref{fig:fig8}. Details about such a 2D dust cluster rotation due to the $\vec{E} \times  \vec{B}$  ion drift have been explained in earlier studies\cite{knopkamagneticrotation,vasilmagneticrotation,satorotation,kawrotation}.

 \subsection{Dynamics of the dusty plasma confined by an insulating ring} 
\begin{figure*}
\centering
 \includegraphics[scale=0.86]{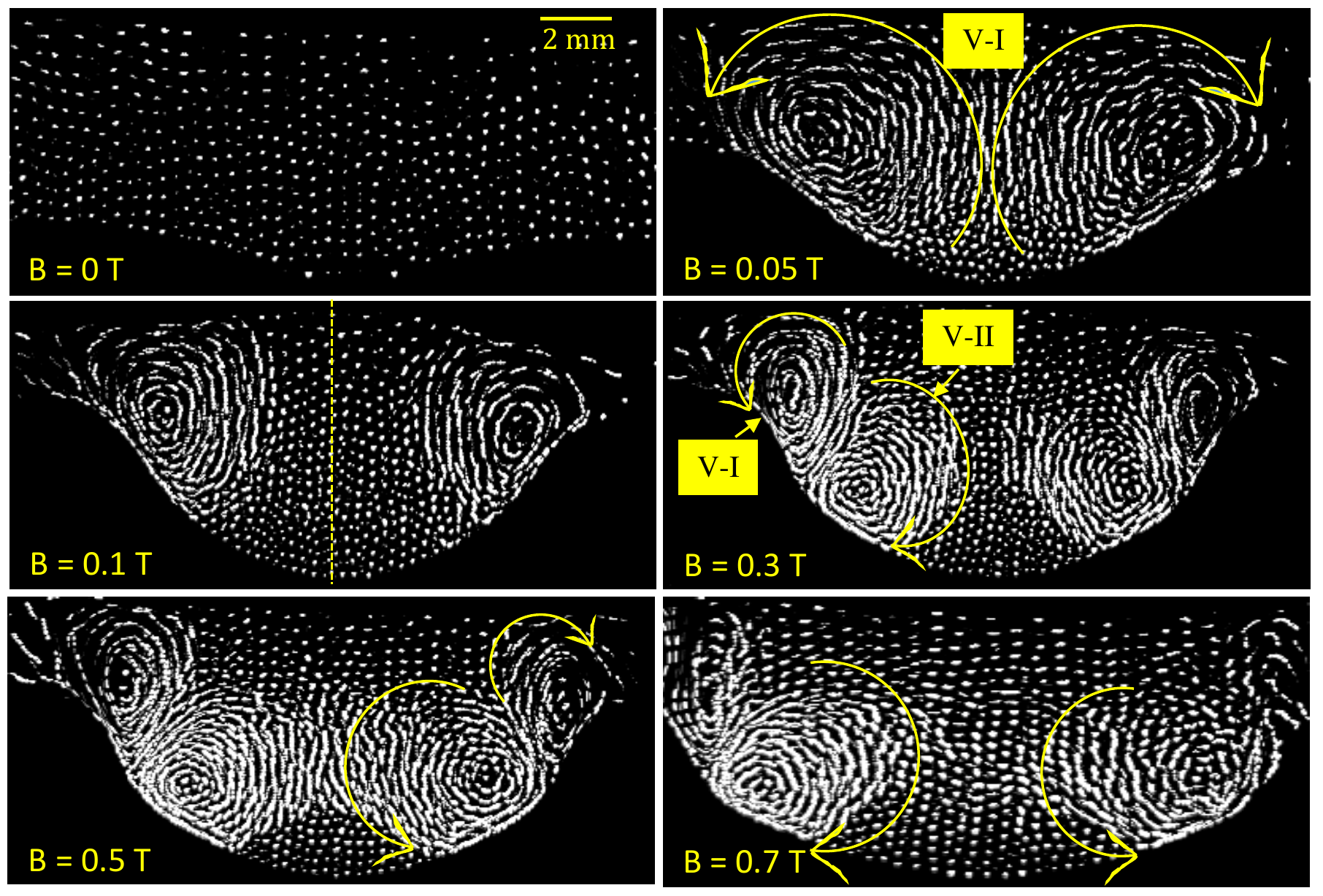}
\caption{\label{fig:fig9} Video images of the dust cloud in the vertical (Y--Z) plane at X = 0 cm. The images at different magnetic fields are obtained by a superposition of five consecutive images at a time interval of 65 ms. A Teflon ring is used to create a potential well at fixed input rf power, P = 3.5 W and argon pressure, p = 35 Pa. The yellow solid line with an arrow indicates the direction of vortex flow in the vertical plane of the 3D dusty plasma.  V--I and V--II represent the edge vortex and central region vortex, respectively.}
\end{figure*}

\begin{figure*}
\centering
 \includegraphics[scale=0.52]{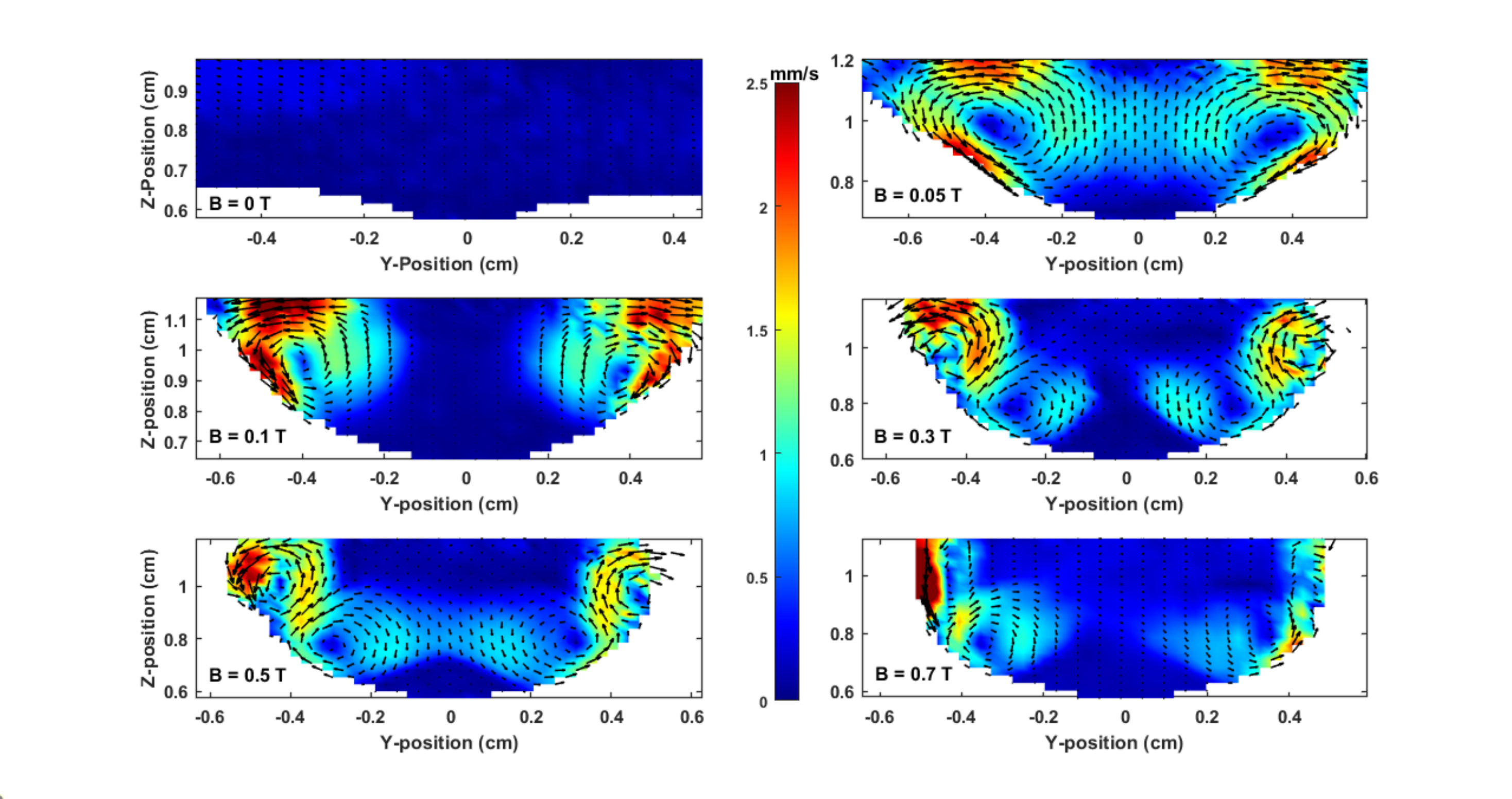}
\caption{\label{fig:fig10} PIV images of the corresponding video images of Fig.~\ref{fig:fig9} at different magnetic field strengths in the vertical (Y--Z) plane. These images are constructed after averaging the velocity vectors over 50 frames. Arrows indicate the direction of rotation in a vortex and color bars represent the magnitude of the velocity of the rotating particles.}
\end{figure*}

\begin{figure*}
 \centering
\subfloat{{\includegraphics[scale=0.35]{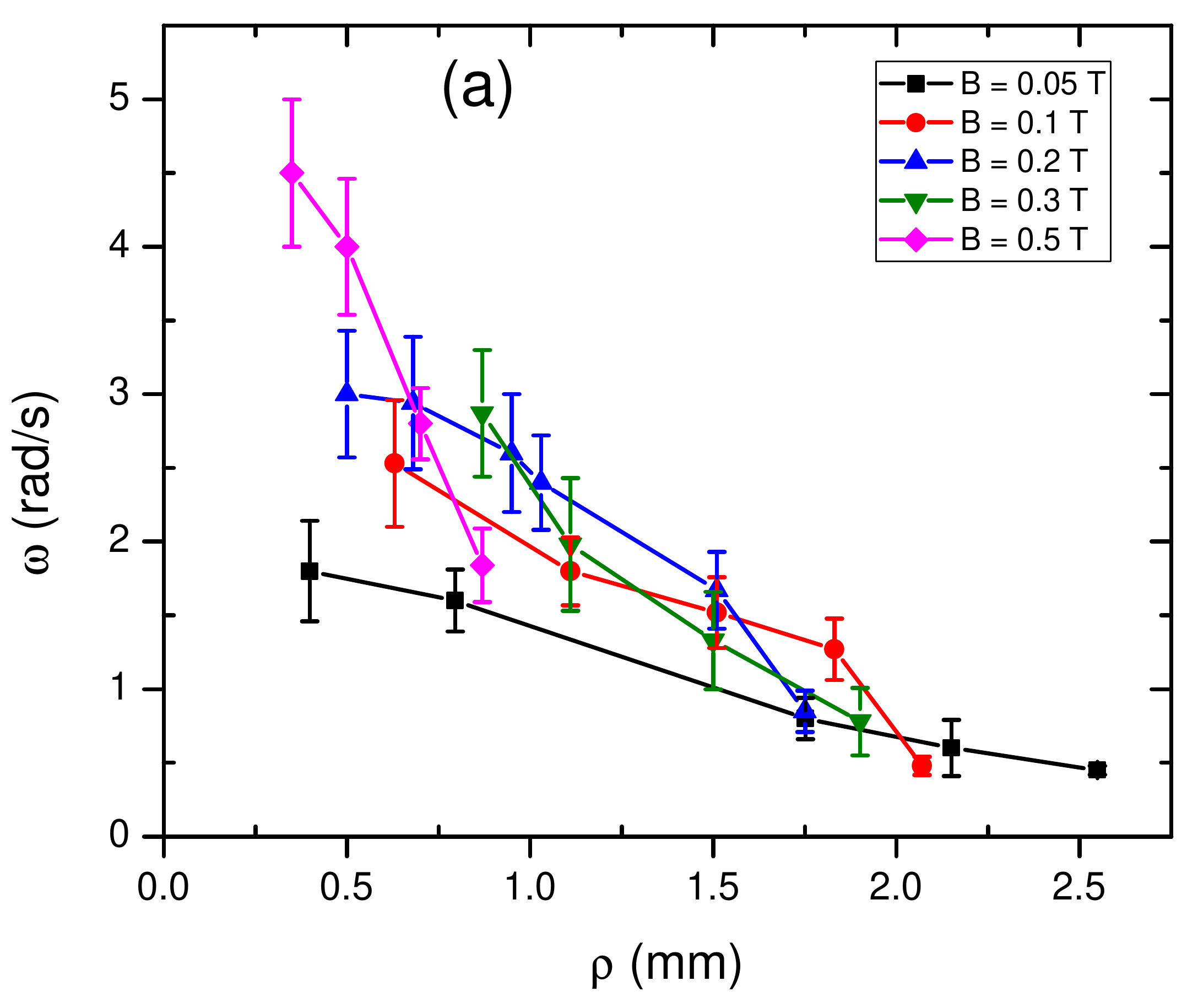}}}%
\qquad
\subfloat{{\includegraphics[scale=0.35]{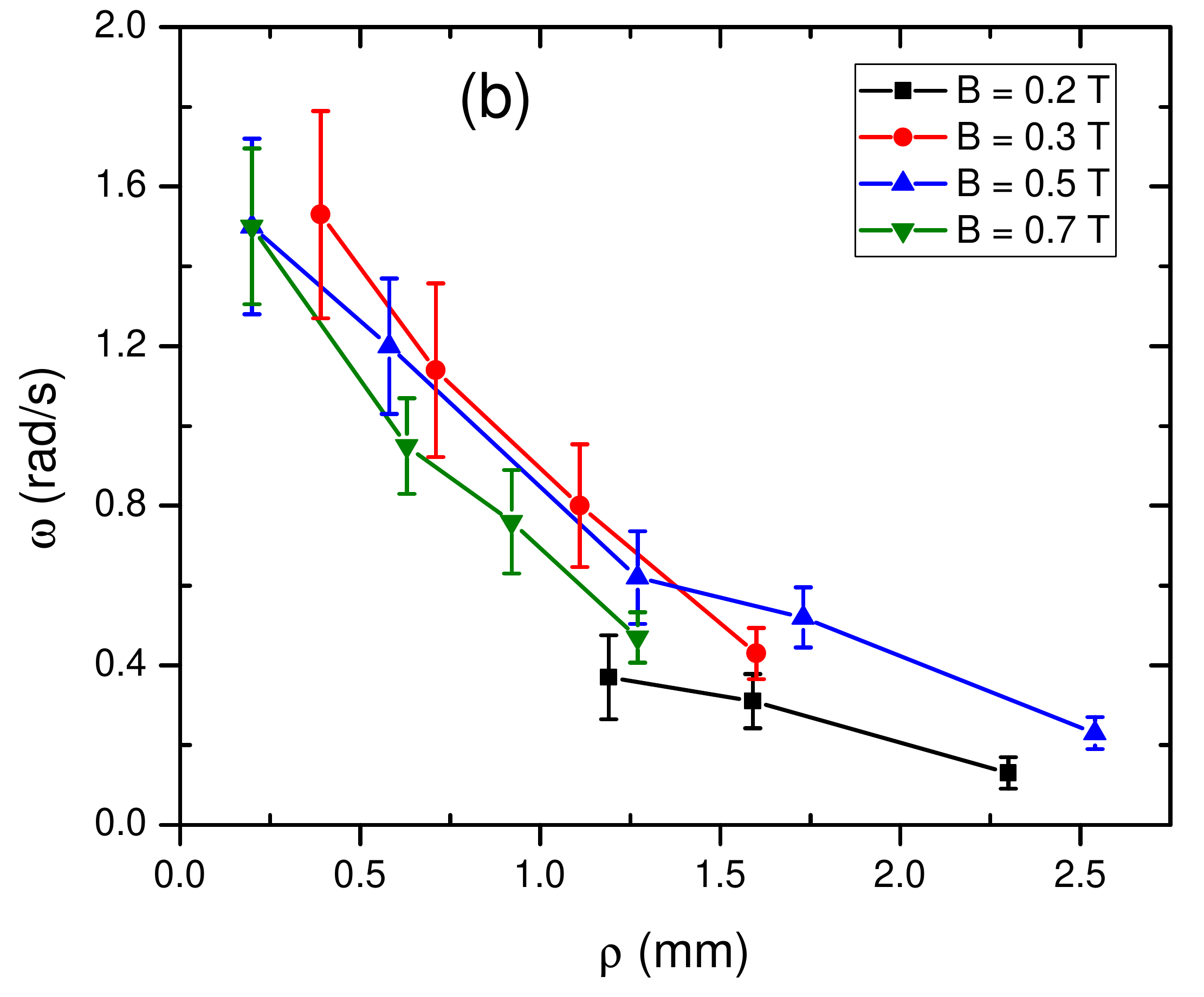}}}
 \caption{\label{fig:fig11} The average angular frequency variation of particles in (a) edge vortex (V--I) and (b) central region vortex (V--II) when the potential well is created by a non-conducting (Teflon) ring at pressure, p = 35 Pa and rf power, P = 3.5 W. }
 \end{figure*}

 In the previous subsection, dust grains were confined in a deep potential well of an aluminium ring, which has the same potential as the lower electrode. In the deep potential well, we have seen mixed motions of weakly and strongly coupled dust grains. Does the dusty plasma show a similar kind of motion when they are confined in a shallow potential well? Is it possible to get both phases of dusty plasma in shallow potential well? To answer these questions, a set of experiments is carried out in a dusty plasma which is confined by a non-conducting (Teflon) ring. Fig.~\ref{fig:fig9} shows the dynamics of a 3D dusty plasma created at argon pressure of p = 35 Pa and input rf power of P = 3.5 W in the presence of an external magnetic field. This figure is obtained after superimposing five consecutive images for tracking the motion of the dust particles. It is observed that the dust grains do not show any directed motion and exhibit only a thermal motion in the shallow potential in the absence of an external magnetic field (B = 0 T). The dust grains also exhibit only a thermal motion at low magnetic field (B $<$ 0.03 T) and after that they start to rotate in this plane and form a pair of counter-rotating vortex structures at B $\sim$ 0.05 T. With increasing magnetic field up to 0.1 T, we observe two well separated vortex structures at the edge of the potential well in this plane. Since the dusty plasma is symmetric about the X-axis in the Y--Z plane, we focus only on the left side region of the dust grain medium. The central region particles have a thermal as well as  $\vec{E_r} \times  \vec{B}$  motion in the X--Y plane at low B. With increasing the magnetic field strength (B $>$ 0.15 T), the dust grains in the central region start to rotate along with the edge vortex in opposite direction,  resulting in a pair of counter-rotating vortex structures. Further increase in the magnetic field from 0.2 T to 0.7 T modifies the characteristics of this pair of counter-rotating vortices. At B $\sim$ 0.7 T, the central region vortex is shifted to the edge of the potential well and a well-separated pair of vortices is observed. At strong magnetic field (B $>$ 0.7 T), the dusty plasma volume reduces and only a few edge particles exhibit a vortex motion. As discussed before, the vortex structure in the vertical plane is a cross-section of the rotating dust torus. Hence at lower magnetic field a single rotating dust torus and a pair of counter-rotating dust tori at higher B is observed.\\
The PIV images of the dusty plasma in the vertical plane at different strengths of the magnetic field (of Fig.~\ref{fig:fig9}) are depicted in Fig.~\ref{fig:fig10}. The earlier discussed PIV analysis technique is used to obtain the angular frequency distribution of rotating particles in a vortex structure at different B-field. The average angular frequency of dust grains in the edge vortex (V--I) and central region vortex (V--II) is presented in Fig.~\ref{fig:fig11}(a) and Fig.~\ref{fig:fig11}(b), respectively. It is clear from Fig.~\ref{fig:fig11} that the angular frequency $\omega$ decreases from the center to the outer edge of the vortices, which is a signature of the differential rotation. At given location of the vortex structure, the value of $\omega$ is observed to be higher at high magnetic field, showing the increase of $\omega$ with increasing external magnetic field. It is also clear from Fig.~\ref{fig:fig11} that particles in the edge vortex always have a higher value of $\omega$ as compared to the central vortex at  given B-field.. Moreover, the particles confined in the potential well of the conducting ring have a higher angular frequency than in the case of the non-conducting ring.  
\section{Discussion} \label{sec:discussion}
The rotating dust grain medium in the external magnetic field can be described by the Navier-Stokes equation with the continuity equation of an in-compressible fluid of constant mass density \cite{vortexmicrogravity, satoprl}
\begin{equation}
 \frac{d \vec{v_d}}{dt} =  -\frac{1}{\rho_d} \nabla p_e+
 \eta\nabla^2 \vec{v_d} + \frac{ q_d}{m_d} (\vec{v_d}  \times  \vec{B}) + 
\frac{1}{m_d} (\vec{F}), 
\end{equation} where
\begin{equation}
\vec{F} = \vec{F_E} + \vec{F_g} + \vec{F_i} + \vec{F_n} + \vec{F_{th}}
\end{equation}
where $\rho_d = m_d  n_d$ is dust mass density, $n_d$ is the dust density, $p_e$ is an electrostatic pressure of the dust grain medium, $m_d$ is mass of the dust grain, $\eta$ is the kinetic viscosity of the dust grain medium, and $\vec{v_d}$ is the dust velocity. In the second equation, $\vec{F_E}$, $\vec{F_g}$, $\vec{F_i}$ , $\vec{F_n}$, $\vec{F_{th}}$ are the electrostatic force, gravitational force, ion-drag force, neutral drag force and thermophoretic force, respectively. The present set of experiments are carried out at an argon pressure of 35 Pa and the dust density $n_d$ varies between $10^4$ to $10^5$ $cm^{-3}$ for given range of B-field (B $<$ 0.8 T). Since dust density or dusty plasma volume at given B-field does not change during the vortex motion, dust grain medium is considered as an in-compressible fluid. Moreover, the rotational velocity of dust grains increases linearly ($>$ 10 \%) with reducing the neutral pressure to 30 Pa at given discharge conditions. It is known that the viscous force depends on the Coulomb interaction (or $Q_d$) among the flowing dust grains \cite{fortovviscosity2}. We observe a slight change ($<$ 3\%) in the ion current during the reduction of pressure  (p = 35 to 30 Pa) at B = 0.05 T (measured with probe), which indicates a negligible change of $Q_d$ or viscosity during the reduction of pressure from 35 to 30 Pa. In PIV images, we see a a slight  velocity gradient, which also favors a less viscous force compared to neutral drag force. Therefore, dust viscous force ($\eta\nabla^2 \vec{v_d}$) is neglected in comparison to neutral drag force ($F_n$) causing dissipation losses during the steady vortex motion. For given discharge conditions, the Lorentz force on the dust grain is of the order of $10^{-18}$ to $10^{-20}$ N which is negligible compared to the external force $F$ of the order of  $10^{-12}$ to $10^{-13}$ N. It shows that the magnetic field does not directly exert a force (or Lorentz force) on charged dust grains to rotate in this plane. The vortex motion in the dust grain medium can be described by the vorticity equation, which can be deduced from equation (1) after taking the curl of both sides \cite{vortexmicrogravity}
\begin{equation}
\frac{d \vec{\omega}}{dt} =  \nabla \times \Big[-\frac{1}{\rho_d} \nabla p_e + \frac{1}{m_d} (\vec{F_E} + \vec{F_g} + \vec{F_i} + \vec{F_n} + \vec{F_{th}})\Big],
\end{equation}
 where $\omega = \nabla \times \vec{v_d}$ is the vorticity of the rotating dust grain medium.
Since the input power is low (P = 3.5 W), the role of thremophoretic force to the rotational motion is considered to be negligible.  $\nabla \times \nabla$$p_e$  and  $\nabla \times \vec{F_g}$  are zero, hence, equation (3) reduces to the form
\begin{equation}
\frac{d \vec{\omega}}{dt} = \frac{1}{m_d} \Big (\nabla \times \vec{F_E} +  \nabla \times \vec{F_i} +  \nabla \times \vec{F_n} \Big)
\end{equation}
For the stationary two-dimensional (2D) vortices in an incompressible dusty plasma, equation (4) can be expressed as \cite{vortexmicrogravity}
\begin{equation}
\nabla \times \vec{F_E} +  \nabla \times \vec{F_i} +  \nabla \times \vec{F_n} = 0
\end{equation}

The electrostatic force on a dust grain of charge $Q_d$ in the electric field of the sheath is given as $\vec{F_E} = Q_d \vec{E}$. Since ions are streaming in the direction of electric field, the ion drag force can be written as $\vec{F_i} = F_i \hat{E}$. Here $F_i$ represents the magnitude of the ion drag force on the dust grain. In the present set of experiments, there is no directed gas flow inside the chamber. Thus, neutrals provide a frictional background to resist the motion of the dust grains. According to Epstein friction \cite{neutraldragepstein}, the neutral friction experienced by the dust grains is $\vec{F}_n = -m_d \nu_{dn} \vec{v}_d$. Here, $\nu_{dn}$ is the dust--neutral friction frequency. After substituting the values of these forces in equation (5), we get the following expression
\begin{equation}
\nabla Q_d \times \vec{E} +  \nabla F_i(E) \times \hat{E} =  m_d \nu_{dn} \nabla \times \vec{v_d}
\end{equation}
In this equation, the L.H.S terms are the energy source for driving the vortex motion and the R.H.S term corresponds to the energy loss during the vortex motion. Thus, for maintaining a stationary vortex flow in the dusty plasma, the energy gain by particles must be balanced by the energy loss. It is clear from equation (5) that the charge gradient and ion drag gradient along with the electric field are possible energy sources to drive the vortex motion. The charge gradient arises due to the inhomogeneity in the background plasma of the dust grain medium \cite{mangirotationpop,vaulinaselfoscillation}. For the ion drag force on the dust grain we take \cite{kharpakiondragforce}
\begin{equation}
\vec{F_i} = \frac{8}{3} \sqrt{2 \pi k_B T_i M_i} r_d^2 n_i  \Big [ 1 + \frac{z \tau}{2} + \frac{z^2 \tau^2 \Lambda}{4} \Big]  (\vec{v_i}-\vec{v_d}) 
\end{equation}
where $M_i$, $T_i$, and $n_i$ denote the mass, temperature and density of ions, respectively. $\vec{v_i} >> \vec{v_d}$ is the average ion velocity (ion drift velocity), $k_B$ is the Boltzmann constant, $z = Z_d e^2 /r_d T_e$ is the dimensionless charge of the dust particle, where $Z_d$ is the dimensionless grain potential in units of $T_e/e$, $\tau = T_e/T_i$ is the electron-to-ion temperature ratio, and $\Lambda$ is the modified Coulomb logarithm integrated over the ion velocity distribution function \cite{kharpakiondragforce}.
It is clear from equation (7) that the ion density and velocity are two major variables to determine the magnitude of ion drag force ($F_i$), hence, the ion drag gradient ($ \nabla F_i(E)$) depends on $n_i$ and $v_i$ in the sheath region of the ring electrode. \par
\begin{figure*}
 \centering
\subfloat{{\includegraphics[scale=0.315]{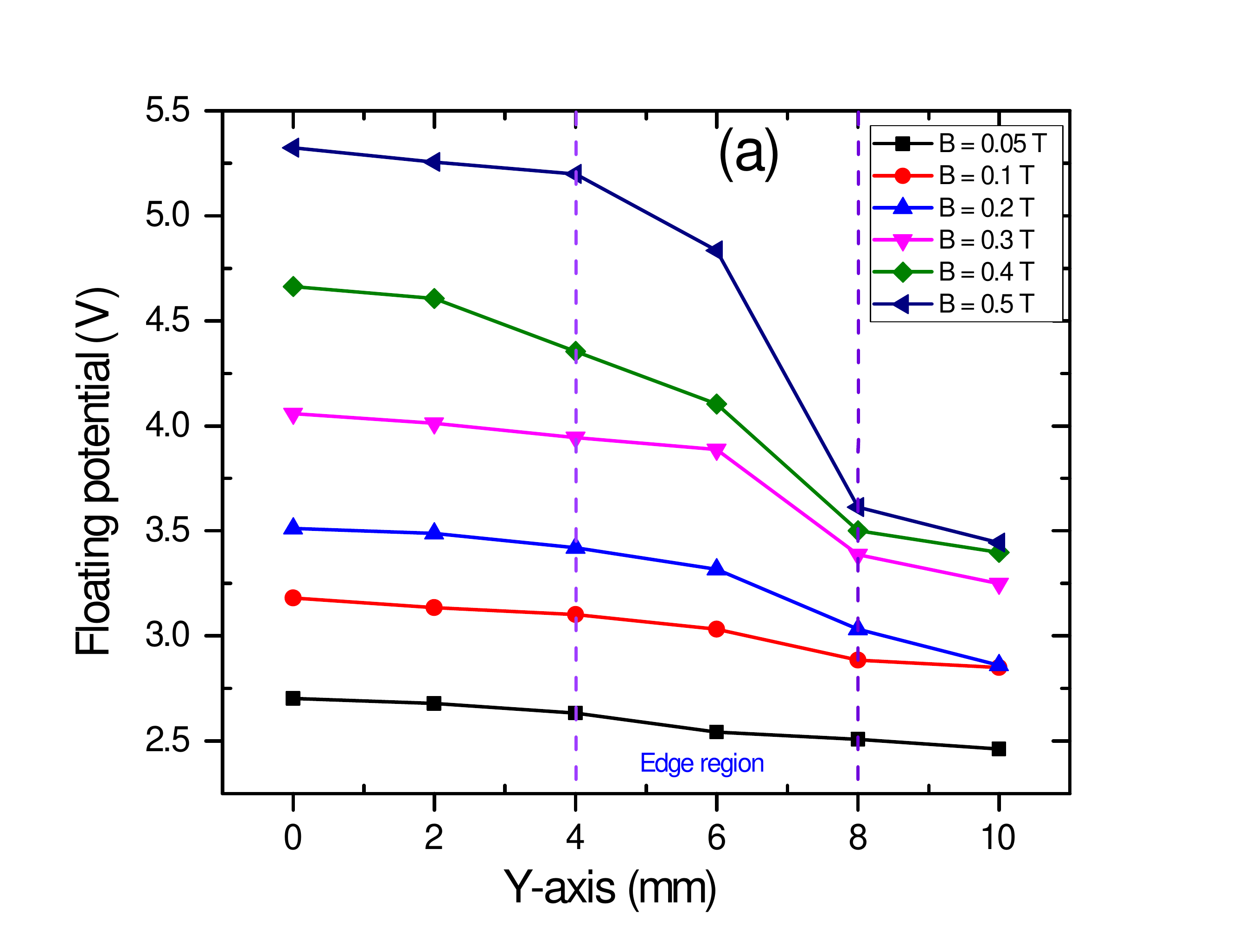}}}%
\subfloat{{\includegraphics[scale=0.3150]{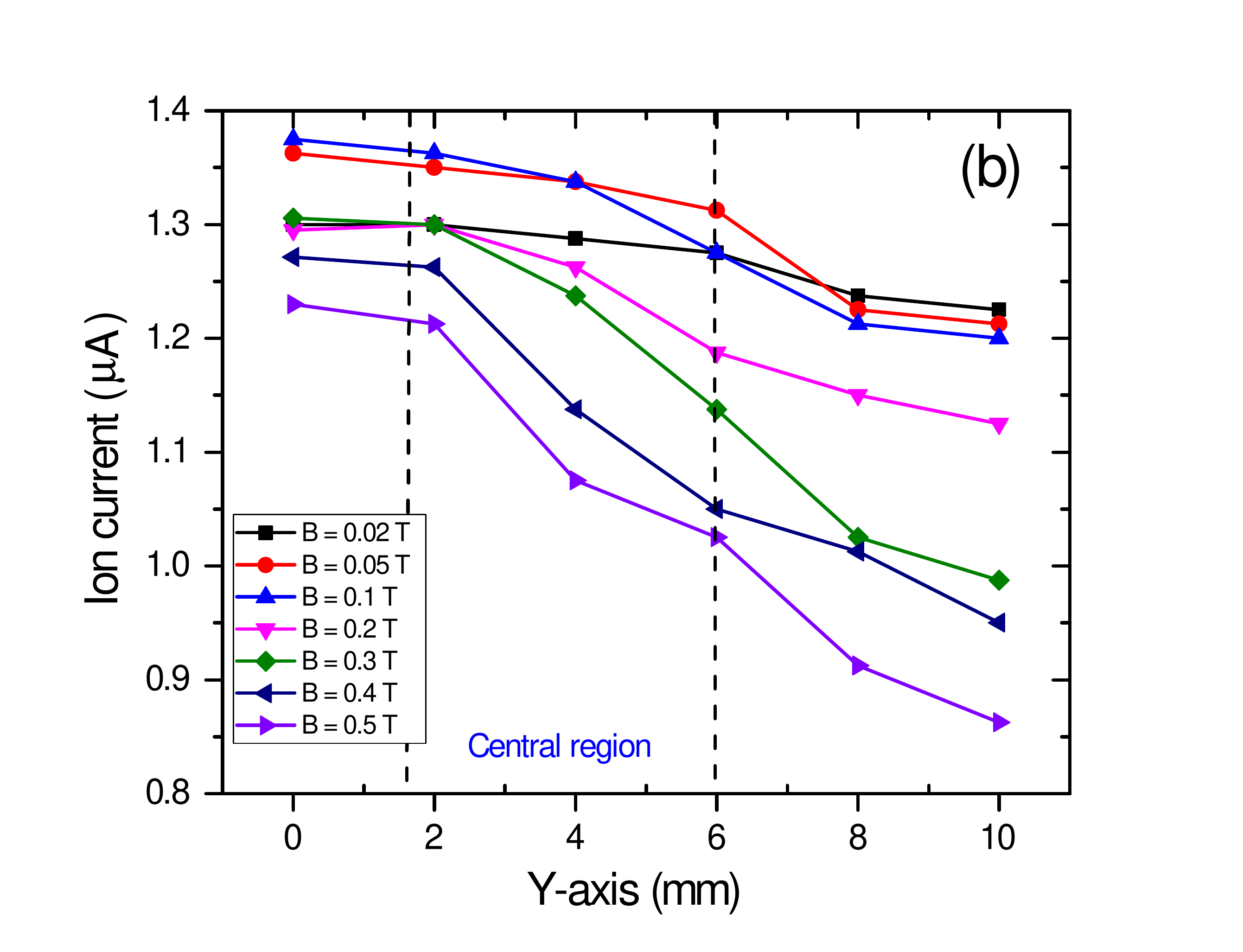}}}
 \caption{\label{fig:fig12}(a) Floating potential measured by a cylindrical probe with respect to ground along the Y-axis (or radial direction) from the center of aluminium ring (Y = 0 mm) to the ring surface (Y = 10 mm) at Z $\sim$ 1.0 cm for various strengths of magnetic fields. (b) Ion saturation current variation at same probe location for different values of B-field. The probe is kept at a fixed bias voltage of -30 V and the ion current is measured for power of P = 3.5 W at a pressure, p = 35 Pa.}
 \end{figure*}

For the quantitative descriptions of dust dynamics in an external magnetic field, it is required to estimate both driving force terms of Eq. 6. In the present case, a single Langmuir probe exerts an electrostatic force to the confined dust grain medium and expels the particles from the confined potential well, which does not allow us to measure the background plasma parameters of the dust grain medium. Another issue to use the single Langmuir probe in moderately collisional magnetized plasma (B $>$ 0.1 T) is the reduction of ion current and formation of secondary plasma near the tip of probe when bias voltage crosses the plasma potential. Therefore, erroneous plasma parameters are expected at higher B-field. It is possible to use the single Langmuir probe in weakly magnetized plasma (B $<$ 0.1 T). However, the spatial variation of plasma parameters (from Y $\sim$ 8 mm to Y = 0 mm) is $<$ 10 \%. The Langmuir probe method which estimates the plasma parameters ($n$, $T_e$ and $V_p$) with an error of $>$ 10 \% in a low density plasma does not give the true values of the plasma parameters at different spatial locations (Y-values). It is known that plasma potential and plasma density can be estimated using the known values of floating potential ($V_f$) and ion saturation current  ($I_i$), respectively  for a given $T_e$ \cite{probemerlino}.
Since $V_f$ and $I_i$ are direct measurable plasma parameters \cite{mangilalpsst}, they are measured along the Y-axis (or radial direction) to understand the dynamics of the dust grain medium qualitatively in presence of the external magnetic field. A cylindrical probe of length $\sim$ 2 mm and radius of 0.125 mm  is used to measure $I_i$ and $V_f$ along the Y-axis (see Fig.~~\ref{fig:fig2} (b)). Fig.\ref{fig:fig12} shows the spatial variation (along Y-axis) of  $V_f$  and $I_i$ at Z $\sim$ 1 cm for a given discharge conditions (P = 3.5 W, p = 35 Pa) at different strengths of B-field. The errors in the measured values of $V_f$ and $I_i$ are within $\pm$ 5\% for the given B-field. The spatial measurement error (along Y-axis) in $V_f$ and $I_i$ is $\pm$ 0.2 mm.\par
It should be noted that dust grains absorb free electrons and reduce the density in dusty plasma. The Havnes parameter \cite{havnes}, $P_h$ = $Z_d n_d/n_i$, decides the density of free electrons in a dusty plasma. It has been assumed that plasma parameters are not strongly affected by the dust grains if $P_h < 1$ and plasma parameters without dust can be used to understand the dynamics of dusty plasma. In the present work, $P_h$ has the value $<$ 0.5 for $n_d$ $\sim$ 3$\times 10^{4}$ to 5 $\times 10^{4}$ $cm^{-3}$, $n_i$ $\sim$ 6$\times 10^{8}$ $cm^{-3}$, and $Z_d \sim$ 5$\times$ $10^3$ in the absence of B-field. The earlier experimental work also suggest that the spatial dependence of the plasma parameters remains nearly same with and without dust grains with a slightly higher or lower value at given location. Therefore, we have measured the ion saturation current and floating potential without dust particles and expect a similar spatial trend in the dusty plasma.\par
In the absence of magnetic field, the dust grains exhibit a thermal motion in the potential well (Fig.~\ref{fig:fig2}). The ion density and velocity gradient, which determine the magnitude of the ion drag gradient ($\nabla F_i(E)$) near the edge of the ring (or edge of the potential well) are not sufficient to drive the vortex flow. With the application of a magnetic field, the confinement of electrons increases the plasma density as well as $T_e$ at low B (B $<$ 0.05 T) \cite{mangilalpsst}. Therefore, the variation in ion current at given location in the plasma is maximum in this range of B-field. In the present set of experiments, the expected variation of $T_e$ along the Y-axis (Y = 0 to 8 mm) is negligible because of the small spatial variation of the plasma density. In our previous study, it has been confirmed that $T_e$ remains almost constant while the plasma density is increased \cite{mangilalpsst}. Therefore, spatial variation of plasma density and plasma potential is considered to be equivalent to the variation of $I_i$ and $V_f$ \cite{probemerlino} as shown in Fig.~\ref{fig:fig12}.\par
\begin{figure}
\centering
 \includegraphics[scale = 0.9]{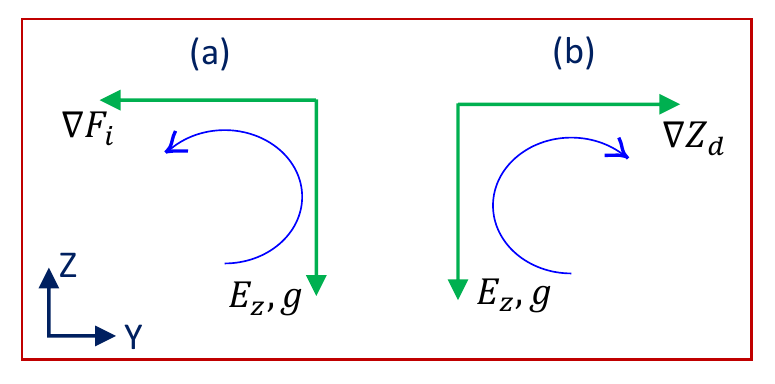}
\caption{\label{fig:fig13}Schematic representation of the rotational direction of particles caused by driving forces (a)  ion shear flow and  (b) dust charge gradient.}
\end{figure}

In the presence of magnetic field, the time-averaged electric potential distribution in the sheath region of the ring is modified which increases the radial electric field ($E_r$) in the sheath region of the ring \cite{2drotationmagneticfield2}. It is clear from Fig.~\ref{fig:fig12}(a) that the gradient of $V_f$ (or $V_p$) in the edge region increases with increasing B-field, which indicates the increase of radial E-field. The velocity of drifting ions ($\vec{v_i} = \mu_i \vec{E_r}$) increases with increasing the strength of B-field. However, the ion density is less in the edge region than in the central region (see Fig.~\ref{fig:fig12}(b)) which points the ion density gradient in the opposite direction than the radial E-field. But the $V_f$ gradient is higher than the ion density gradient in edge region, which indicates that the role of velocity gradient (or radial E-field) is stronger than ion density to drive the vortex flow through radial ion shear flow or gradient (Eq. 6 and 7) in the presence of the electric field ($E_z$) against the dissipation losses by neutrals\cite{vortexmicrogravity,icedustyvortex,laishramshearflow}. Hence the rotating edge particles form a vortex structure (V--I) in the Y--Z plane or a single rotating dust torus in 3D dusty plasma. The gradient of $V_f$ in the edge region continuously increases as the strength of B-field is increased, which shows higher radial E-field at strong B. In our observations, we see a reduction in size of the edge vortex (see Fig.~\ref{fig:fig4}) which also confirms the role of radial E-field in the formation of edge vortex (V--I). A schematic diagram of the direction of rotation due to the shear in the ion drag force along with electric field is shown in Fig.~\ref{fig:fig13}(a). In our experiments, the direction of the edge vortex flow (see Fig.~\ref{fig:fig3} and Fig.~\ref{fig:fig9}) is found to be similar to that predicted by theoretical model which confirms the role of the ion drag gradient or ion shear flow along with electric field for creating the edge vortex (V--I) in a vertical plane or a rotating dust torus in 3D dusty plasma. The magnitude of the driving force ($\vert\nabla F_i(E) \times \vec{E_z} \vert$) determines the angular frequency of the rotating particles in the dust torus, which strongly depends on the potential distribution in the edge region (or radial electric field). Since a B-field increases the gradient of potential in the edge region (see Fig.~\ref{fig:fig12}(a)), $\omega$ increases with increasing the strength of the external magnetic field. \\
The radial component of  the electric field ($E_r$) in the presence of the magnetic field also drives an azimuthal flow of the ions through the $\vec{E_r} \times  \vec{B}$ drift. These drifted ions in the azimuthal direction transfer their momentum to the dust particles and set them into rotational motion \cite{2dclusterrotation,knopkamagneticrotation,satorotation}. Due to the  $\vec{E_r} \times  \vec{B}$ drift motion of ions, the rotating dust grain in a vortex can move to the next vertical plane. This is one of possible causes to have the azimuthal velocity component along with the radial velocity component in the X--Y plane of a rotating dust torus (see Fig.~\ref{fig:fig7}). In the case of the conducting ring electrode, particles which are confined in the deep sheath region or bottom of the potential well are assumed to be less strongly coupled. We assume that these particles are either a result of agglomeration of particles or the presence of some massive impurity particles in the sample of mono-dispersive MF particles. They do not participate in the vortex motion leading to a dust torus. These particles rotate in the azimuthal direction in the X--Y plane, which can be seen in Fig.~\ref{fig:fig8}.\par
At the higher magnetic field, B $>$ 0.15 T, dust grains in the central region rotate in opposite direction to the edge vortex flow and form a pair of counter-rotating dust vortices (or dust tori). It is seen in Fig.~\ref{fig:fig12}(a) that the gradient of $V_f$ is stronger in the edge region than in the central region of plasma at higher B-field. Fig.~\ref{fig:fig12}(b) shows that the ion density gradient increases in central region of plasma at higher strength of B-field. However, the role of $v_i$  ($\vec{v_i} = \mu_i \vec{E_r}$) determining the magnitude of ion drag gradient dominates over ion density $n_i$ in the central region dust grains medium. The $\vec{E_r} \times  \vec{B}$  ion drift induced motion of dust grains  (Fig.~\ref{fig:fig8}) at given B-field also indicates the radial outward flow of ions. Both results are in favour of a negligible role of ion density gradient to drive the vortex flow in opposite direction (V--II). Thus, the second driving force (Eq. 6) which arises due to the dust charge gradient may excite the central region vortex motion  (V--II)  in the presence of E-field or gravity \cite{vaulinajetp,vaulinaselfoscillation, selfexcitedmotion,mangirotationpop,mangilalvortexseries,horizontalrotation}. The spatial plasma inhomogeneity is one of the possible causes for the dust charge gradient \cite{mangilalvortexseries}. Figure~\ref{fig:fig12}(b) confirms the inhomogeneity in the plasma density in the central region at higher B. It has been proven by a theoretical model that even 1 \% of dust charge gradient can initiate the vortex motion through an instability \cite{vaulinaselfoscillation,vaulinasripta2004}. The direction of rotation due to the dust charge gradient along with E-field or gravity is shown in Fig.~\ref{fig:fig13}(b). At B $>$ 0.15 T, ions start to be magnetized and electrons are fully magnetized. The magnetic field reduces the plasma loss to the ring electrode, hence, the density of the plasma confined in the ring electrode increases. In spite of the ion current, the plasma glow intensity around and in the ring electrode (without dust particles) noticed with naked eyes also confirms the plasma density enhancement at higher B (B $>$ 0.15 T). It is fact that plasma density increases with increasing the glow intensity in rf discharge.  Therefore, we expect the plasma density gradient towards the center of ring at strong B-field.
\par
In our recent study, the role of the B-field on the surface potential (or charge) on the spherical probe (or large dust grain) has been studied \cite{mangilalpsst}. The grain charge is determined by the balance of electrons and ions flux to its surface \cite{Charging}. It has been observed in experiments that the dust charge strongly depends on the plasma density \cite{mangilalpsst} in the magnetized plasma.  It collects more electrons in the higher density region than that in low density region in the presence of B-field. A higher electron current makes its surface more negative. A reduction of the electron flux to dust surface in the low plasma density region makes its surface less negative \cite{floatingpotential,mangilalpsst}. Since the plasma is quasi-neutral in the central region, the electron density is almost equal to the ion density. In the present set of experiments, the plasma density (or electron density) gradient in the central region (expected $>$ 10 \%) creates a dust charge gradient in the same direction. This dust charge gradient $\nabla Q_d = e\nabla Z_d$, orthogonal to the gravitational force ($\vec{F}_g$), or ion drag force ($\vec{F}_I$) acting on the dust particles drives the vortex flow \cite{vaulinajetp,vaulinaselfoscillation, selfexcitedmotion,mangirotationpop,mangilalvortexseries,horizontalrotation}. With increasing magnetic field from 0.2 T to 0.8 T, the angular frequency of the rotating particles in the central region vortex (V--II) increases. This increase in $\omega$ is due to the increase in the magnitude of the dust charge gradient which depends on the gradient of the plasma density. Hence the angular frequency of the central vortex flow (V--II) increases with increasing the strength of the external B-field. The direction of the vortex flow is in the direction of the charge gradient, which is in accordance with the available theoretical model \cite{vaulinaselfoscillation}.\\
\section{Summary} \label{sec:summary}
The present work highlights the collective dynamics of a 3D dusty plasma in the presence of an external strong magnetic field. The 3D dusty plasma is created in a capacitively coupled discharge by placing an additional conducting or non-conducting ring of a specific diameter and width on the lower powered electrode. The observed results of the reported work are summarized as:
\begin{enumerate}
\item A 3D dusty plasma in a strong magnetic field can be created using an additional conducting or non-conducting ring of a particular inner diameter and width (or thickness) in an rf discharge.
\item The magnetic field modifies the potential distribution of the edge region or sheath region and a potential gradient is develop near the edge region of ring, which gives rise to a radial electric field, hence, increases the ion shear flow. The ion drag gradient along with the E-field drives the vortex flow near the edge of potential well and an edge vortex (V--I) is formed.
\item At higher magnetic fields (B $>$ 0.15 T), electrons are fully magnetized and ions start to get magnetized, which increases the plasma density in the central region of the dusty plasma with increasing B-field. Thus, the magnetic field creates an inhomogeneous plasma in the confining potential well for B $>$ 0.15 T. The plasma inhomogeneity is responsible for the dust charge gradient, which along with the electric field or gravity drives the central vortex motion.
\item The angular frequency has a radial dependence which suggests a differential or sheared rotational motion in the vortices.
\item The magnetic field enhances the angular frequency of the rotating particles in the vortex. Particles in the potential well of the conducting ring have a higher angular frequency than that of the non-conducting ring.
\end{enumerate}  

The formation of the edge vortex is discussed on the basis of a radial ion shear flow or ion drag gradient together with the electric field in the presence of an external magnetic field. Weakly coupled particles rotate in the azimuthal direction due to the $\vec{E_r} \times  \vec{B}$ drift of the ions in this plane. The onset of the vortex motion of the central region particles at higher magnetic field is understood by the presence of a dust charge gradient along with the gravity or electric field. The direction of the vortex motion based on a theoretical model also supports our qualitatively discussed sources of driving force.\par
The present work confirms that dynamics of 3-dimensional dusty plasma is determined by the magnitude of radial ions shear flow near the confining boundary, resulting in a vortex motion in the vertical (Y--Z) plane. In a strong B-field, the dust charge gradient due to plasma inhomogeneity also induces the vortex motion in the 3D strongly coupled dusty plasma. However, a 2-dimensional dust cluster in a similar confining potential well always rotates in the azimuthal or  $\vec{E_r} \times  \vec{B}$ direction even at strong B-field. There are many challenges to diagnose the dusty plasma in the presence of strong magnetic field using the electrostatic probes as well as spectroscopy. For getting a better understanding of the collective dynamics of strongly magnetized 3D dusty plasma, computer simulation for such plasma systems are necessary. In the future, we plan to perform numerical simulations to understand the presented experimental results in more details.
\section{Data Availability Statement} The data that support the findings of this study are available from the corresponding author upon reasonable request.
\bibliography{aipsamp}
\end{document}